\documentclass[
	aps, prd, reprint,
	10pt, notitlepage, a4paper,
        floats, floatfix,
	amsmath, amssymb, amsfonts, eqsecnum,
	superscriptaddress,
	showpacs, showkeys,
	nofootinbib,
 	longbibliography,
]{revtex4-1}

\usepackage{graphicx}
\usepackage[usenames,dvipsnames]{xcolor}
\usepackage{tikz}
\usepackage{xspace} 
\usepackage{dcolumn}
\usepackage{bm}
\usepackage{multirow} 
\usepackage[utf8]{inputenc} 
\usepackage{latexsym}
\usepackage{float}
\usepackage{ulem}
\renewcommand{\emph}[1]{\textit{#1}}
\usepackage{longtable}
\usepackage{times}

\xdefinecolor{mylinkcolor}{rgb}{0,0,0.5}
\usepackage[
	bookmarksnumbered, bookmarksopen, bookmarksopenlevel=2,
	breaklinks=true,
	colorlinks=true, filecolor=mylinkcolor, citecolor=mylinkcolor,
	linkcolor=mylinkcolor, urlcolor=mylinkcolor, menucolor=mylinkcolor,
]{hyperref}

\newcommand{\AEI}{\affiliation{Max Planck Institute for Gravitational Physics (Albert Einstein Institute), Am M\"uhlenberg 1, Potsdam 14476, Germany}}
\newcommand{\Maryland}{\affiliation{Department of Physics, University of Maryland, College Park, MD 20742, USA}}

\newcommand{\CENTRA}{\affiliation{Centro Multidisciplinar de Astrof\'isica --- CENTRA, Departamento de F\'isica,
 	Instituto Superior T\'ecnico --- IST, Universidade de Lisboa --- ULisboa,
 	Avenida Rovisco Pais 1, 1049-001 Lisboa, Portugal}}

\def\be{\begin{equation}}
\def\ee{\end{equation}}
\def\bea{\begin{eqnarray}}
\def\eea{\end{eqnarray}}
\newcommand{\bes}{\begin{subequations}}
\newcommand{\ees}{\end{subequations}}

\def\vct#1{{\bm{#1}}}

\def\nl{\\ & \quad}
\def\nnl{\nonumber \\ & \quad}

\def\PM{pm}

\DeclareMathOperator{\Order}{\mathcal{O}}

\allowdisplaybreaks

\begin{document}

\title{Dynamical Tides in General Relativity:\\Effective Action and Effective-One-Body Hamiltonian}
\hypersetup{pdftitle={Dynamical tides in General Relativity: Effective Action and Effective-One-Body Hamiltonian}}

\author{Jan Steinhoff}
\AEI \CENTRA

\author{Tanja Hinderer}
\Maryland \AEI

\author{Alessandra Buonanno}
\AEI \Maryland

\author{Andrea Taracchini}
\AEI

\date{\today}

\begin{abstract}
Tidal effects have an important impact on the late inspiral of compact binary systems containing
neutron stars. Most current models of tidal deformations of neutron stars assume that the
tidal bulge is directly related to the tidal field generated by the companion, with a constant response
coefficient. However, if the orbital motion approaches a resonance with one of the internal
modes of the neutron star, this adiabatic description of tidal effects starts to break down, and the
tides become dynamical. In this paper, we consider dynamical tides in general relativity due to the
quadrupolar fundamental oscillation mode of a neutron star. We devise a description of the effects
of the neutron star's finite size on the orbital dynamics based on an effective point-particle action
augmented by dynamical quadrupolar degrees of freedom. We analyze the post-Newtonian and test-particle
approximations of this model and incorporate the results into an effective-one-body Hamiltonian.
This enables us to extend the description of dynamical tides over the entire inspiral. We demonstrate
that dynamical tides give a significant enhancement of matter effects compared to adiabatic tides,
at least for neutron stars with large radii and for low mass-ratio systems, and should therefore be
included in accurate models for gravitational-wave data analysis.
\end{abstract}


\maketitle

\section{Overview}
The much anticipated era of gravitational-wave astronomy recently began
with the observation of gravitational waves from binary black-hole mergers
by Advanced LIGO \cite{Abbott:2016blz,Abbott:2016nmj}.
Still the two LIGO detectors \cite{TheLIGOScientific:2014jea} have not
reached design sensitivity yet, and will be augmented by Advanced Virgo
\cite{TheVirgo:2014hva}, KAGRA \cite{Aso:2013eba}, and
LIGO-India \cite{LIGOIndia} in the future.
Such a network of ground-based gravitational-wave observatories is needed for
improving the sky localization of sources and thus enable targeted electromagnetic follow-up
observations. This is a particularly fascinating
prospect for neutron stars in compact binary coalescences where the merger or disruption
is expected to generate for instance short gamma-ray bursts \cite{Fernandez:2015use}.

Maximizing the science gains from gravitational-wave observations requires accurate models of the binary dynamics as matched-filtering templates for data analysis. Of particular importance for
the analytic description of the dynamics of a neutron star in a binary is a detailed model for tidal interactions. The purpose of the present paper is to develop
a model for dynamical tides in general relativity and to incorporate it
into the effective-one-body (EOB) formalism \cite{Buonanno:1998gg,Buonanno:2000ef},
which has been providing LIGO and Virgo with waveform models to detect signals, infer their  
astrophysical properties, and test general relativity~\cite{Abbott:2016blz,Abbott:2016nmj,TheLIGOScientific:2016wfe,TheLIGOScientific:2016src,Abbott:2016izl}.

\subsection{Newtonian dynamical tides\label{tideintro}}
It is instructive to review dynamical tidal effects for an irrotational ideal fluid in Newtonian gravity. 
For simplicity, consider an isolated star in an external gravitational field.
The external tidal field deforms the star and displaces its fluid elements away from their
equilibrium position. At linear order in this perturbation,
the displacement of the
fluid elements can be represented as a superposition of normal modes of oscillation, where the coefficients are dynamical
(time dependent) mode amplitudes. The normal mode that dominates the tidal interaction is the quadrupolar fundamental (f-)mode.
The f-modes can be understood as standing waves on the surface of the
star\footnote{By definition, the f-modes have no nodes of oscillation inside the star and the
oscillation amplitude grows towards the surface. Their overtones are called p-modes.} that
are efficiently excited through tidal forces. Resonances between the orbital motion and the quadrupolar f-mode in Newtonian
gravity were first discussed for ordinary stars by Cowling \cite{Cowling:1941}
and much later for neutron stars
\cite{Reisenegger:1994, Shibata:1993qc, Kokkotas:1995xe, Lai:1993di, Ho:1998hq, Lai:2006pr}. 
However, these studies in Newtonian gravity are of limited applicability to physically realistic neutron stars since they are strongly self-gravitating objects. The purpose of the present work is to overcome these limitations and develop a rigorous model for dynamical tidal excitations in general relativity.

The quadrupolar oscillations of a neutron star due to the f-mode can be described by a dynamical quadrupole $Q^{ij}$, with $i,j=1,2,3$, obeying the equation of motion of a tidally driven harmonic oscillator. We do not include a damping of the oscillator since the neutron-star viscosity is low
and therefore the star is not tidally locked \cite{Kochanek:1992wk,Bildsten:1992my}.
The corresponding Lagrangian is \cite{Flanagan:2007ix}
\begin{equation}\label{LNewton}
L_\text{DT} = \frac{1}{4 \lambda \omega_f^2} \left[ \dot{Q}^{ij} \dot{Q}^{ij}
  - \omega_f^2 Q^{ij} Q^{ij} \right] - \frac{1}{2} E_{ij} Q^{ij} ,
\end{equation}
where a dot denotes a time derivative, the numerical constant $\omega_f$ is the angular frequency of the f-mode,
$\lambda$ is the tidal deformability which is related to the Love number \cite{Love:1909}, and $E_{ij}$ is the quadrupolar tidal field.
In terms of the Newtonian gravitational
potential $\Phi$ the tidal field is $E_{ij} = \partial_i \partial_j \Phi$. The Lagrangian $L_\text{DT}$, together with a point-mass action, can be used as a model for a neutron star
in a binary, supplemented by the usual action of the Newtonian gravitational field.
A generalization of Eq.~(\ref{LNewton}) to additional modes is straightforward.

The meaning of the tidal deformability is best understood in
the limit of adiabatic tides, which is given by $\omega_f \rightarrow \infty$ for our normalization of $L_\text{DT}$.
In this limit, the kinetic term in the Lagrangian (\ref{LNewton}) drops out and a variation 
of $Q^{ij}$ leads to $Q^{ij} = - \lambda E^{ij}$. That is, the quadrupole instantaneously
follows the external tidal field $E^{ij}$ with the proportionality factor being the
tidal deformability $\lambda$. For finite $\omega_f$, one can consider an equilibrium
solution of the oscillator as in Ref.~\cite{Flanagan:2007ix} and as we discuss in Appendix \ref{equilibrium}.
This solution can be used to determine initial conditions for the quadrupole equations of motion.

\begin{figure}
\centering \includegraphics[width=0.47\textwidth]{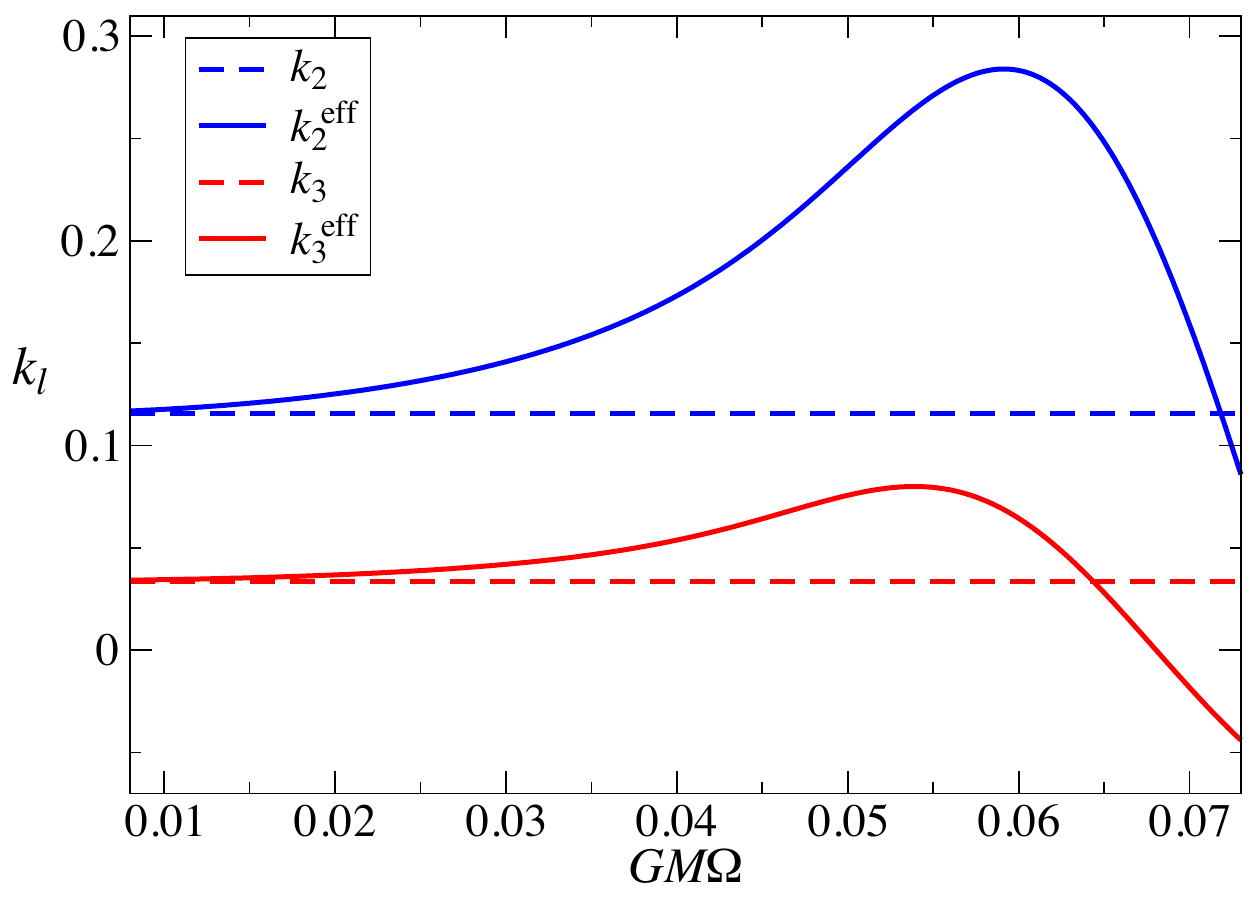}
\caption{Dimensionless effective tidal deformability from a two timescale approximation under
leading-order radiation reaction [see Sec.~\ref{keff} with the replacement $r= (GM) (GM \Omega)^{-2/3}$, in units with $c=1$ and for an H4 equation of state and mass $1.35M_\odot$].
The index $l$ refers to the multipolar order, such that $k_2$ is the quadrupolar dimensionless
tidal deformability and $k_3$ is the octupolar one. \label{keffintro}}
\end{figure}
To characterize the effects of dynamical tides we introduce an effective tidal deformability
$\lambda_\text{eff}$ that depends on the binary separation. Since the separation evolves under gravitational
radiation reaction, $\lambda_\text{eff}$ is in fact a function of time.
We define $\lambda_\text{eff}$ through
\begin{equation}\label{leffdef}
\lambda_\text{eff} = - \frac{E_{ij} Q^{ij}}{E_{kl} E_{kl}} .
\end{equation}
Note that in the adiabatic case $\lambda_\text{eff} = \lambda$.
When we evaluate Eq.~(\ref{leffdef}) for
an inspiral using a dynamical quadrupole, the function $\lambda_\text{eff}$ can be
understood as a varying tidal deformability.  The deviation of $\lambda_\text{eff}$ from its
constant value $\lambda$ is an indication of the impact
of dynamical tides.

In Sec.~\ref{keff} we derive an approximate analytic expression for
$\lambda_\text{eff}$ using a
two timescale method. The result is shown in Fig.~\ref{keffintro} and
displays the enhancement of tidal effects due to dynamical tides close to merger or disruption.
The quantities shown in this figure are the dimensionless Love numbers which are related to the deformability by
\begin{equation}
\label{k2}
k_{\ell} = \frac{(2 \ell - 1)!!}{2} \frac{G \lambda_{\ell}}{R^{2 \ell+1}} ,
\end{equation}
where $R$ is the radius of the neutron star and $\ell$ is the multipolar
order ($\ell = 2$ for the quadrupole considered here, i.e., $\lambda \equiv \lambda_2$).
We work in units where $c=1$, but we keep Newton's constant $G$. We use greek letters to denote spacetime indices that run over $\{0,1,2,3\}$ and latin letters running over the values $\{1,2,3\}$ for 3-dimensional spatial components.

\subsection{Qualitative expectations for relativistic effects in dynamical tides\label{framedrag}}
Relativistic corrections to the Newtonian tidal interactions discussed above are important to accurately describe tidal effects of 
binary neutron stars. Such corrections were computed in Ref.~\cite{Vines:2010ca} within a post-Newtonian (PN) approximation to 1PN order and applicable for any kind of tides, and for the case of adiabatic tides the 2PN order was calculated in Ref.~\cite{Bini:2012gu}.
These studies showed that relativistic corrections enhance the tidal
force acting on the body, which is a statement on the interaction term $E_{ij}Q^{ij}$ in Eq.~(\ref{LNewton}).
Moreover, by virtue of the equivalence principle, Eq.~(\ref{LNewton}) provides an intuitive
description of the relativistic case in a local freely falling coordinate system attached
to the neutron star. Such local observer experiences a relativistic redshift relative
to an observer at spatial infinity and also a frame dragging due to gravito-magnetic
fields. This has interesting consequences for dynamic tides.

The physical consequence of the redshift effect can be understood as follows.
All frequencies measured in the neutron-star's frame are redshifted from the perspective
of an observer measuring the gravitational waves at spatial infinity. This means that
the f-mode frequency seen by the distant observer is redshifted with
respect to the constant f-mode frequency $\omega_f$ in the rest frame of the neutron star.
Conversely, from the perspective of the neutron star, the frequency of the driving tidal force
is larger compared to that inferred by an asymptotic observer.
This redshift effect is expected to enhance the dynamical tidal effects, since it shifts the resonance
with the f-mode to a lower orbital frequency. The radiation reaction is therefore
smaller at the resonance, such that the system spends more time close to the resonance
and transfers more energy from the orbital motion to the tidal excitation.

\begin{figure}
\centering
\includegraphics{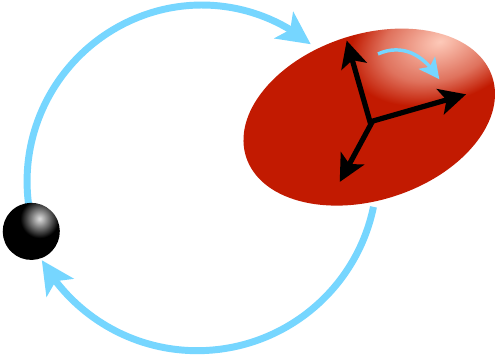}
\caption{The frame of the tidally deformed neutron star is dragged in the
direction of the orbital motion.\label{frameill}}
\end{figure}
The consequence of the frame-dragging effect due to the gravito-magnetic field is somewhat opposite to the redshift
effect. For a comparable mass system, the dominant angular momentum is the orbital
one. Hence the neutron-star frame is dragged in the direction of the orbital motion 
as illustrated in Fig.\ \ref{frameill}.
The orbital frequency in this dragged frame is therefore lower
than for the distant observer. The frame dragging thus effectively shifts
the f-mode to a higher frequency. This is analogous to the Zeeman effect
for the splitting of atomic spectral lines in the presence of a magnetic field. Similarly, a bulge on the star rotates clockwise or counter-clockwise within the orbital plane,
as a free oscillation. Invoking the equivalence principle, one infers that
the bulge rotates with the constant f-mode frequency $\omega_f$ in both directions
in the neutron-star frame. However, this frame is dragged as seen from a distant observer.
This observer therefore sees different frequencies for the clockwise and counter-clockwise
oscillations: the frequency of the bulge traveling in the direction of the orbit
is shifted to larger values, while the frequency of the bulge traveling in the opposite direction is shifted to lower values.
However, since the external tidal field always tracks the orbital motion, only the mode with
the raised frequency is excited. A similar effect also occurs for neutron stars with spin \cite{Ho:1998hq}, where, however, the direction of the dragging depends on the orientation of the spin. For a neutron star with a large spin that is anti-aligned with the orbital angular momentum, the resonance frequency is effectively lowered since in that case the spin drags the
frame in the direction opposite to the tidal force.

The frame dragging is usually encoded in various spin interactions in a Hamiltonian
formulation of the binary dynamics. This
is true also for the frame dragging acting on the dynamical tides. Noether's theorem applied to the
rotational invariance of Eq.~(\ref{LNewton}) shows that the tides contribute to the
total angular momentum through a ``tidal spin'' given by the antisymmetric tensor
$S_Q^{ij} = 2 Q^{k[i} \dot{Q}^{j]k}/(\lambda \omega_f^2)$. To obtain a complete tidal 
model it is essential to include a covariant generalization of this spin 
in place of the ordinary relativistic spin interaction terms in the Hamiltonian,
whose importance was alluded to in Ref.~\cite{Vines:2010ca}, and
which becomes obvious from Eq.~(\ref{LFD}) below.

\subsection{Action for relativistic dynamic tides}
Dynamical tides in general relativity have been studied in
the case of a test-mass orbiting a neutron star
\cite{Kojima:1987tk, Ruoff:2000et, Gualtieri:2001cm, Pons:2001xs} and for comparable masses in the PN limit focusing on r-modes~\cite{Flanagan:2006sb}. Resonances due to tidal interactions have also been seen in numerical-relativity simulations \cite{Gold:2011df} for binaries on eccentric orbits. An interesting dynamical response to a \emph{stationary} tidal field was found recently
for a slowly rotating neutron star \cite{Landry:2015snx}.
The authors of Refs.~\cite{Maselli:2012zq, Ferrari:2011as, Maselli:2013mva} have developed a dynamical model for the tidal interaction of neutron stars by approximating them as triaxial ellipsoids with self-similar internal isodensity surfaces. This model takes into account
the strong self-gravity of the neutron star, but does not include mode resonances in an explicit way.
The effect on the gravitational-wave phase was found to be negligible \cite{Maselli:2012zq}.
We come to a different conclusion here when dynamical tides are allowed to become resonant.

Let us write down a 4-dimensional covariant and minimally coupled form of the
Lagrangian (\ref{LNewton}) as
\begin{equation}\label{LQ}
L_\text{DT} = \frac{z}{4 \lambda} \left[ \frac{1}{z^2 \omega_f^2} \frac{D Q_{\mu\nu}}{d \sigma} \frac{D Q^{\mu\nu}}{d \sigma}
  - Q_{\mu\nu} Q^{\mu\nu} \right] - \frac{z}{2} E_{\mu\nu} Q^{\mu\nu} ,
\end{equation}
with the full action principle of the matter being 
\begin{equation}
\label{SP}
S = \int d\sigma \, L , \qquad
L = - m z + L_\text{DT} ,
\end{equation}
where $D$ denotes a covariant parameter derivative,
$z = \sqrt{- u^\mu u_\mu}$, $u^\mu = \dot{y}^\mu$, and the worldline of the particle is $y^\mu(\sigma)$
with $\sigma$ being a generic worldline parameter.
The signature of spacetime is $+2$.
Note that in this notation $z^2 = - u_\mu u^\mu$, and the factors of $z$ are introduced such that the action is invariant under reparametrizations
of the worldline parameter $\sigma$. For the gauge choice of $\sigma$ adopted later on,
$z$ takes on the physical meaning of the redshift factor.
The 4-dimensional tidal field $E_{\mu\nu}$ is the electric part of the Weyl tensor
$C_{\mu\nu\alpha\beta}$ given by
\begin{equation}\label{defE}
E_{\mu\nu} = C_{\mu\alpha\nu\beta} \frac{u^\alpha u^\beta}{z^2} ,
\end{equation}
which is reparametrization invariant and is a symmetric-tracefree spatial tensor in the rest frame,
i.e., $E_{\mu\nu} u^\nu = 0$, $E_{[\mu\nu]} = 0$, and $E^\mu{}_{\mu} = 0$. Similarly, the 4-dimensional
quadrupole tensor $Q^{\mu\nu}$ is required to be a symmetric-tracefree spatial tensor in the rest-frame,
\begin{gather}
Q_{\mu\nu} u^{\nu} = 0 , \label{Qconstraint} \\
Q^{[\mu\nu]} = 0, \quad Q^\mu{}_{\mu} = 0 . \label{QSTF}
\end{gather}
These are covariant constraints that reduce the quadrupole degrees of freedom to the
correct physical ones. We explicitly relate $Q^{\mu\nu}$ to a SO(3) tensor in Sec.\ \ref{impose}
and highlight the connection of the equations of motion
derived from the Lagrangian~(\ref{LQ}) to the dynamics of a generic extended body given by Dixon \cite{Dixon:1979} in Sec.\ \ref{theory} B.  In Sec.\ \ref{approx} we compute the Lagrangian (\ref{LQ}) within the PN approximation for the orbital dynamics. The PN results agree with the 1PN tidal Lagrangian derived in Ref.~\cite{Vines:2010ca}. However, the formalism developed in this paper features several advances beyond the standard PN approach such as (i) elucidating the role of the frame effects discussed above, which emerge from the constraint on the quadrupole in Eq.~(\ref{Qconstraint}) and the covariant derivative in Eq.~(\ref{LQ}), (ii) exhibiting the redshift factors explicitly, and (iii) revealing a direct mapping between tidal effects and known PN results for spinning bodies, which we explain in Sec.\ \ref{approx}.

\subsection{Body and orbital zones\label{bodyzone}}
The link between the action~(\ref{SP}) describing a point particle with a dynamical quadrupole and the actual extended neutron star is established by introducing various zones in which different approximation schemes are valid. For instance, in the PN
approximation, one introduces a body zone for each object where gravity can
be strong, an orbital zone (or near zone) where the PN expansion in weak gravitational fields and
slow motion can be applied, and a radiation zone where the emitted gravitational
waves are weak and propagate with the speed of light.

The connection between the zones can be rigorously established using matched asymptotic
expansions as summarized in Ref.~\cite{Blanchet:2013haa}.
For binary black holes, an explicit construction of all zones has been developed in the context of initial data for numerical-relativity simulations~\cite{Yunes:2005nn,Yunes:2006iw,JohnsonMcDaniel:2009dq,Gallouin:2012kb,Mundim:2013vca,Ireland:2015cjj}.
For neutron stars, the process of matching between body and orbital zones encodes the tidal interactions. 
An explicit construction of all the zones analogous to that for black holes is not yet available. However, this does not prevent us from obtaining a complete description of the orbital dynamics, since this requires only knowledge of the body's multipole moments~\cite{Flanagan:1997fn,Racine:2004hs}. For stars with low compactness such as white dwarfs, the matching calculations can also be done by applying the PN approximation to the interior of the star, which was worked out to 1PN order by Damour, Soffel, and Xu \cite{Damour:1990pi, Damour:1991yw, Damour:1992qi, Damour:1993zn}.

The matching of the body and orbital zones can be achieved by using
a point-particle action as an intermediary,
since it provides an immediate physical understanding, like the harmonic
oscillator action in Eq.~(\ref{LQ}). Once the parameters defined by the action ($\lambda$ and
$\omega_f$) are fixed through some matching, one can apply the point-particle
model to a PN description of the orbital dynamics. One can think of
the body zone being effectively shrunk to a point. Conversely, from the perspective
of one of the bodies, the orbital scale can be expanded to spatial infinity.
This leaves an isolated body in an external field, which is a rather simple
setting in which the parameters in the action can be matched. For instance, the tidal
parameters $\lambda$ and $\omega_f$ can be obtained from linear perturbations
of a spherically symmetric relativistic star. This approach properly incorporates the strong gravity inside
relativistic stars, which is reflected in the numerical values for
$\lambda$ and $\omega_f$. The quadrupolar Love number $\lambda$ was first obtained from linear
perturbations of a relativistic star in
Ref.~\cite{Hinderer:2007mb} and generalized to higher multipoles in Refs.~\cite{Binnington:2009bb,Damour:2009vw}. The latter study also raised important subtleties in defining the Love numbers through
such a matching procedure \cite{Damour:2009vw}. Subsequently
Ref.~\cite{Kol:2011vg} showed how these subtleties are avoided in the case of
nonrotating black holes. The rotating case is not settled, but progress
has been made in the slow rotation approximation \cite{Poisson:2014gka, Landry:2015zfa, Pani:2015hfa, Pani:2015nua}.
The matching of the f-mode frequency is likewise a
delicate problem and the frequency entering the action~(\ref{LQ}) is
distinct from the complex quasi-normal mode frequencies
\cite{Kokkotas:1999bd, Nollert:1999ji}. We discuss all these issues in detail in  Sec.~\ref{GRaction}.

\subsection{Effective-one-body Hamiltonian}
The impact of dynamical tides over adiabatic ones is expected to be
noticeable only close to the f-mode resonance. This occurs in the
strong-field regime of general relativity, where the PN approximation
loses accuracy. Dynamical tides in general relativity therefore require a
method which is applicable to the nonlinear orbital regime, such as numerical relativity.
However, to enable the generation of a large bank of gravitational-wave templates for data analysis, a computationally much less expensive
approach is needed. The EOB model is currently used
for this purpose since it provides an accurate description of the
entire gravitational-wave signal by combining analytical
information from PN and black-hole
perturbation theory into a single framework \cite{Buonanno:1998gg,Buonanno:2000ef}.
The accuracy of the model has been further improved through a calibration to numerical relativity
\cite{Taracchini:2013rva,Nagar:2015xqa}, thus creating a synergy of the most powerful
tools to describe relativistic compact binaries.

The EOB model was extended to tidal effects in 
Refs.~\cite{Damour:2009wj,Baiotti:2011am,Bernuzzi:2012ci,Bini:2012gu,Bini:2014zxa,Bernuzzi:2014owa},
but restricted to adiabatic tides. The purpose of the present
paper is to improve the description of matter effects by considering dynamical
tidal effects in the EOB Hamiltonian. In contrast to Ref.~\cite{Bini:2012gu}, our construction implements the test-particle
results without introducing poles in the Hamiltonian (see Secs.~\ref{qcirc} and \ref{EOBLR}).
This is important for neutron-star--black-hole systems, where for certain mass ratios 
the poles might be reached during the final stages of the binary evolution. The main result for the EOB Hamiltonian is given by 
Eqs.\ (\ref{HEOB}), (\ref{Heff2})--(\ref{muDT}), (\ref{Ecirc})--(\ref{zcirc}), and (\ref{Zcirc2})
for circular orbits. This result is accurate to 1PN order and further contains partial information at 2PN order in Eq.~(\ref{Ecirc}) determined by matching to the adiabatic limit from Ref.~\cite{Damour:2009wj}.
We also study different, structurally less motivated, implementations of dynamical tides
in the EOB Hamiltonian to verify that our conclusions are not an artifact of the specific
implementation. Our results are the foundation of EOB waveforms with fully dynamical tides that have been compared against numerical-relativity
simulations in Refs.~\cite{Hinderer:2016eia, inprep2}. The tidal EOB Hamiltonian can equivalently be obtained from the generic 1PN tidal Lagragian derived in Ref.~\cite{Vines:2010ca}. The benefit of starting from a relativistic action is that it leads to immediate insights into the structure of the terms, thus providing physical intuition as well as useful guidance for devising an EOB resummation of tidal effects.

\vspace{0.3cm}
The plan of this paper is the following. We first discuss the general relativistic point-particle
action encoding dynamical tides in Sec.\ \ref{theory}. To express the terms in this action explicitly, we specialize to the PN and test-particle approximations in Sec.\ \ref{approx}.
This is the basis for the EOB Hamiltonian derived in Sec.\ \ref{DTEOB}, following the
construction principles outlined in Sec.\ \ref{construction} and making use of the
gauge freedom from Sec.\ \ref{gauge}. Finally, the results are discussed in
Sec.\ \ref{discuss} where we compare waveforms including dynamical tides
with waveforms using only adiabatic tides. We find that dynamical tides are an important
physical effect for certain realistic nuclear equations of state and mass ratios.

\section{Theory of relativistic dynamical tides\label{theory}}
In this section we discuss in detail the effective point-particle action for dynamical tidal effects in
general relativity. We first review Newtonian dynamic tides to motivate the covariant form of the relativistic action~(\ref{LQ}) which we determine within an effective-field-theory approach. Next, we consider the equations of motion and Legendre transformations that bring the action into a convenient form. Lastly, we impose the constraints by separating the time and spatial components of the tidal variables to derive an action that involves only the physical tidal degrees of freedom.

\subsection{The effective action\label{GRaction}}
Below we discuss the reasoning that led us to posit the particular form of the Lagrangian (\ref{LQ}) for a  relativistic action that describes quadrupolar mode oscillations of a deformable body. We start by reviewing the Newtonian description of stellar oscillations to make the relation between the mode amplitudes and the quadrupole degrees of freedom $Q_{ij}$ explicit. Subsequently, we use the effective-field-theory approach for compact binaries developed by Goldberger and Rothstein ~\cite{Goldberger:2004jt,Goldberger:2007hy} to obtain a covariant version of the Newtonian action that leads to Eq.~(\ref{LQ}). Previous work on this topic already derived the  quadrupolar interaction terms~\cite{Goldberger:2005cd, Goldberger:2009qd} and considered a dynamical quadrupole in the context of absorption from the black-hole horizon~\cite{Goldberger:2005cd}.
Other work~\cite{Bini:2012gu} obtained an effective action in the limit of an expansion around the adiabatic case. Here, we go beyond these studies by deriving a general effective action for a fully dynamical quadrupole that describes mode oscillations of a deformable body. We further discuss subtleties related to the identification of the coupling constants $\lambda$ and $\omega_f$, survey additional terms that could in principle contribute to the action, and argue that in the case of interest here these terms are negligibly small.

Generic tidal perturbations of a Newtonian star can be decomposed into its normal modes of oscillation~\cite{Chandrasekhar:1964}, and are an extensively studied topic. An action principle for the mode amplitudes was formulated by Alexander \cite{Alexander:1987}, and also derived from
Lagrangians for an ideal fluid polytrope \cite{Gingold:1980}, for homentropic stars \cite{Rathore:2002si}, and from an effective-field-theory approach \cite{Chakrabarti:2013xza}. These action principles rely on treating the amplitude of each mode as a harmonic oscillator.
Since the unperturbed star is rotationally
symmetric, the modes fall into irreducible representations of SO(3). This implies that the quadrupolar mode
variables, which are usually decomposed into a spherical-harmonic basis with $l=2$ and $m=-l,\cdots,l$, can equivalently be described by rank-two symmetric-tracefree tensors, denoted here by $A^{ij}$, with $A^{[ij]} = 0 = A^{ii}$.
The Lagrangian for the quadrupolar f-mode amplitudes $A^{ij}_f$ therefore has the form
\begin{equation}\label{LDTf}
L_\text{DT} = \frac{1}{2} \dot{A}^{ij}_f \dot{A}^{ij}_f - \frac{\omega_f^2}{2} A^{ij}_f A^{ij}_f - \frac{I_f}{2} E_{ij} A^{ij}_f + \dots ,
\end{equation}
where the constants $\omega_f$ and $I_f$ are the angular frequency and coupling constant, also known as the ``overlap integral"~\cite{Press:1977}, of the mode, $E_{ij}$ is the quadrupolar tidal field, and the dots denote possible nonlinear interaction terms.
The Lagrangian in Eq.~(\ref{LDTf}) differs from Eq.~(\ref{LNewton}) only by a choice of normalization, where 
\begin{align}
A^{ij}_f = \frac{1}{I_f} Q^{ij} , \qquad
\lambda = \frac{I_f^2}{2 \omega_f^2} .
\end{align}
It is straightforward to extended this result to several quadrupolar modes by adding copies of Eq.~(\ref{LDTf}) for each mode. However, if the normal-mode expansion fails to represent the complete solution for the perturbed star, copies of Eq.~(\ref{LDTf}) for each mode will be insufficient to represent the entire quadrupolar response of the star and additional terms of the form $E_{ij} E_{ij}$ must be included in the Lagrangian~(\ref{LDTf}) to compensate for the residual discrepancy. For Newtonian perfect fluid stars, the normal modes are complete~\cite{Cox:1980} and hence no such additional terms are required. In this case, the constants $\omega_f$ and $I_f$ (or $\lambda$) entering the Lagrangian are easily identified with quantities computed from linear perturbations of a fluid star ~\cite{Cox:1980, Love:1909}. The dominant modes for tidal interactions are the f-modes, whose tidal coupling constants $I_f$ are several orders of magnitude larger than those of other quadrupolar modes \cite{Shibata:1993qc, Kokkotas:1995xe}, hence we neglect those other modes here.

To obtain a relativistic generalization of the Newtonian Lagrangian (\ref{LNewton}) we employ the effective-field-theory approach to the gravitational interaction
of compact objects~\cite{Goldberger:2004jt}. In this approach, the interaction terms in the action are determined by writing down all possible operators consistent with the symmetries (general covariance, parity, and time reversal), and redefining variables to eliminate couplings that involve accelerations \cite{Damour:1990jh}.
 For the linear, electric-type, quadrupolar interactions these considerations lead to a single interaction term derived in Ref.~\cite{Goldberger:2005cd} and given by $\sim \int d\sigma E_{\mu\nu}Q^{\mu\nu}$, with the relativistic tidal field $E_{\mu\nu}$ defined in terms of the spacetime curvature in Eq.~(\ref{defE}). This generalizes the Newtonian coupling $\int dt E_{ij}Q^{ij}$ and the Newtonian definition of $E_{ij}$. The remaining steps in mapping from the Newtonian to the relativistic action consist in replacing time derivatives with covariant derivatives along the worldline, and inserting factors of $z$ to ensure invariance of the action under reparametrizations of the parameter $\sigma$. In general, as discussed above in the Newtonian case, tidal couplings of the form $E_{\mu\nu}E^{\mu\nu}$ may need to be added to the Lagrangian (\ref{LQ}). Such terms would account for the incompleteness of modes which is known to occur in
general relativity, as well as for other quadrupolar modes besides the f-modes. However, as in the Newtonian case, the coupling coefficients of these additions are estimated to be small~\cite{Chakrabarti:2013lua} and we therefore neglect these additional terms here.

As mentioned in Sec.~\ref{bodyzone}, the relativistic effective action~(\ref{SP}) discussed above describes the binary only on an orbital scale, where the coefficients $\lambda$ and $\omega_f$ remain undetermined and must be
linked to quantities describing a perturbed relativistic fluid star through a matching procedure. In contrast to the Newtonian case, the relativistic nonlinearities introduce subtleties into this identification and can lead to counter-intuitive results. For instance, the Love number $\lambda$ of black holes vanishes \cite{Kol:2011vg}, which is impossible to reproduce through a superposition of damped
mode amplitudes as would be done when extrapolating Newtonian results. While neutron stars are less compact than black holes, they nevertheless enclose strong gravitational fields and might inherit some non-intuitive features. A rigorous definition of their tidal deformability coefficients $\lambda$ requires performing an analytic continuation in the dimensionality of spacetime as done for the case of black holes in Ref.~\cite{Kol:2011vg} or, as a more practical but less rigorous alternative, using the prescription for neutron stars developed in Ref.~\cite{Chakrabarti:2013lua}. Likewise, the real mode frequency parameter $\omega_f$ in the Lagrangian follows from a matching of the orbital and body zones as discussed in detail in Ref.~\cite{Chakrabarti:2013lua}. The boundary conditions of this matching are different from those used to define the complex quasi-normal mode
frequencies \cite{Kokkotas:1999bd, Nollert:1999ji}, yet the numerical value of $\omega_f$ determined in this way turns out to be very close to the value of the real part of the quasi-normal mode frequency~\cite{Chakrabarti:2013lua}.

Having discussed the construction of the relativistic action for fully dynamical quadrupoles, it is also useful to consider the limiting case far from a resonance where the quadrupole is nearly adiabatic, to establish a connection with previous work in Refs.~\cite{Bini:2012gu,Chakrabarti:2013xza}. The effective-field-theory paradigm states that all degrees of freedom with frequencies above the orbital frequency should be integrated out of the action. Thus, when restricting the description to
tidal driving frequencies that cannot excite the f-mode,  the tidal Lagrangian (\ref{LQ}) is approximated by a quasi-adiabatic Lagrangian~\cite{Bini:2012gu,Chakrabarti:2013xza}
\begin{equation}\label{adiabaticGR}
L_\text{qAT} = \frac{\lambda}{4} E_{\mu\nu} E^{\mu\nu}
+ \frac{\lambda'}{4} \frac{D E_{\mu\nu}}{d \sigma} \frac{D E^{\mu\nu}}{d \sigma} + \dots ,
\end{equation}
with the dots denoting similar terms with higher-order derivatives of $E^{\mu\nu}$. The first term in Eq.~(\ref{adiabaticGR}) corresponds to the adiabatic limit and the second term is the first correction due to dynamical tides, with the coefficient $\lambda^\prime$ determined in terms of ($\lambda$, $\omega_f$) by the Taylor expansion
\begin{equation}\label{lambdaprime}
\frac{\lambda \omega_f^2}{\omega_f^2 - \omega^2} = \lambda + \lambda' \omega^2 +{\cal O}(\omega^4),
\end{equation}
i.e., $\lambda^\prime=\lambda/\omega_f^2$, and similarly for the omitted higher-order terms. Close to the resonance, such an expansion of the Lagrangian around the adiabatic limit in Eq.~(\ref{adiabaticGR}) breaks down since the resonance corresponds
to a pole in the response. Cases for which the inspiral terminates well before the resonance is reached
could be adequately described by retaining a finite number of terms in Eq.~(\ref{adiabaticGR}). This would avoid the introduction of additional dynamical variables for the quadrupole, which is computationally expensive. However, in Sec.~\ref{keff} we introduce a significantly more useful method for reducing the computational
cost while still capturing the nonlinear features of the resonance.

\subsection{Equations of motion}
To study the dynamics described by the action (\ref{LQ}) we first obtain the equations of motion using a manifestly covariant variation as described in detail in Ref.~\cite{Steinhoff:2014kwa}. Ignoring the constraint in Eq.~(\ref{Qconstraint}) for the sake of clarity,\footnote{The constraint (\ref{Qconstraint}) is preserved if the secondary constraint $P^{\mu\nu} u_\mu = 0$ holds. The method of Lagrange multipliers legitimizes our procedure, since it fixes the multipliers of these constraints to zero, up to terms of negligible order in the curvature.} this leads to
\begin{gather}
\frac{D p_\mu}{d \sigma} = \frac{1}{2} S_Q^{\alpha\beta} R_{\alpha\beta\rho\mu} u^\rho - \frac{1}{6} \nabla_\mu R_{\alpha\rho\beta\sigma} J_Q^{\alpha\rho\beta\sigma} , \label{pEOM} \\
\frac{2 \lambda}{z} \frac{D P_{\mu\nu}}{d \sigma} =
- Q_{\mu\nu} - \lambda E_{\mu\nu}. \label{PEOM}
\end{gather}
Here, we have introduced a ``tidal spin'' tensor $S_Q^{\mu\nu}$ associated with the angular momentum of the dynamical quadrupole and a rank-4 quadrupole moment $J_Q^{\mu\nu\alpha\beta}$ given by
\begin{align}
S_Q^{\mu\nu} &= 4 Q^{\rho[\mu} P^{\nu]}{}_\rho , \label{SQ} \\
J_Q^{\alpha\rho\beta\sigma} &=  - \frac{3}{z} u^{[\alpha} Q^{\rho][\beta} u^{\sigma]} . \label{JQ}
\end{align}
The generalized momenta in Eqs.~(\ref{pEOM}) and ~(\ref{PEOM}) are defined by
\begin{align}
p_\mu &= \frac{\partial L}{\partial u^\mu} , \label{pdef} \\
P_{\mu\nu} &= \frac{\partial L}{\partial \left( \frac{D Q^{\mu\nu}}{d \sigma} \right)}
 = \frac{1}{2 \lambda \omega_f^2 z} \frac{D Q_{\mu\nu}}{d \sigma} \label{QEOM},
 \end{align}
where the partial derivatives of the Lagrangian are calculated assuming the functional dependence
\begin{equation}
L = L\left(u^\mu, Q^{\mu\nu}, \frac{D Q^{\mu\nu}}{d \sigma}, R_{\mu\nu\alpha\beta}, g_{\mu\nu} \right) .
\end{equation}
Our convention for the Riemann tensor is
\begin{equation}
R^{\mu}{}_{\nu\alpha\beta} = \Gamma^{\mu}{}_{\nu \beta , \alpha}
	- \Gamma^{\mu}{}_{\nu \alpha , \beta}
	+ \Gamma^{\rho}{}_{\nu \beta} \Gamma^{\mu}{}_{\rho \alpha}
	- \Gamma^{\rho}{}_{\nu \alpha} \Gamma^{\mu}{}_{\rho \beta} ,
\end{equation}
where $\Gamma^{\mu}{}_{\nu \beta}$ is the Christoffel symbol.

Since we are considering here an irrotational matter configuration,
the presence of spin terms in the equations of motion requires further
explanation. The interpretation is that the tidal bulge carries an
angular momentum given by Eq.~(\ref{SQ}) since the bulge points towards the companion and thus travels around the neutron star's surface during an orbit. However, this angular motion of the bulge is due to fluid elements undergoing
only a radial motion; hence the neutron star itself remains irrotational. Yet, it has a net spin given by the sum of the spin due to the rotation of the fluid, which vanishes in the case considered here, and the
tidal angular momentum $S_Q^{\mu\nu}$. The dynamics of the tidal spin $ S_Q^{\mu\nu}$ are analogous to those of an intrinsic spin,
obeying the generic form of the equations of motion for the spin-dipole
found by Dixon~\cite{Dixon:1979},
\begin{equation}\label{SQEOM}
\frac{D S^{\mu\nu}_Q}{d \sigma} = 2 z E^{\rho[\mu} Q^{\nu]}{}_\rho
  = 2 p^{[\mu} u^{\nu]} + \frac{4}{3} R_{\alpha\beta\rho}{}^{[\mu} J_Q^{\nu]\rho\beta\alpha},
\end{equation}
which can be verified using the quadrupolar equations of motion~(\ref{PEOM}) and (\ref{QEOM}). Dixon's general multipolar approximation scheme fully determines the equations of motion only for the linear momentum $p_\mu$ and spin-dipole of the body. The equations of motion for all higher multipoles are not restricted by the conservation of energy and momentum, and depend on the internal structure of the body. Therefore, information about the internal dynamics of the higher multipoles must be supplemented to Eqs.~(\ref{pEOM}) and (\ref{SQEOM}). An example of such supplemental information to complete the set of equations of motion is the oscillator dynamics describing f-modes in Eqs.~(\ref{PEOM}) and (\ref{QEOM}). This example also illustrates that the tensors describing the spin and higher multipole moments in Dixon's equations of motion, in this case the quantities in Eqs.~(\ref{SQ}) and (\ref{JQ}), are in general merely mathematical structures that represent combinations of more fundamental degrees of freedom.

Having developed insights into the covariant dynamics discussed above, we next turn to the idea of using the action (\ref{LQ}) to derive a fully constrained Hamiltonian that can be mapped to an EOB model. This requires transformations of Eq.~(\ref{LQ}) that involve the following steps. First, we replace all velocities in favor of the conjugate momenta, include the mass-shell constraint in the transformed action, and perform a decomposition of all the quantities into time and space directions. Next, we obtain explicit expressions for the various terms in the resulting Lagrangian within the PN approximation, as well as in the test-particle limit, and construct the corresponding Hamiltonian. Finally, we investigate several possibilities for mapping this information onto the EOB model. In the subsequent sections we present a detailed discussion of each step in this procedure.

\subsection{Legendre transformations}
Following Refs.~\cite{Steinhoff:2010zz,Steinhoff:2014kwa}, we apply a Legendre transformation and rewrite the action 
in Eq.~(\ref{LQ}) in the following equivalent form
\begin{equation}\label{Raction}
S = \int d\sigma \left[ P_{\mu\nu} \frac{D Q^{\mu\nu}}{d \sigma} + R_Q \right] ,
\end{equation}
where
\begin{equation}\label{RQ}
R_Q = - m z - z \lambda \omega_f^2 P_{\mu\nu} P^{\mu\nu}
- \frac{z}{4 \lambda} Q_{\mu\nu} Q^{\mu\nu} - \frac{z}{2} E_{\mu\nu} Q^{\mu\nu} .
\end{equation}
The action~(\ref{Raction}) has the advantage that
the complicated covariant derivative of $Q^{\mu\nu}$ appears
only linearly and only in a simple kinematic term, which is convenient for explicit calculations.

A further Legendre transformation can be performed to replace $u^\mu$ by $p_\mu$.
This is interesting since it manifestly brings the action into first-order form in all variables,
which is necessary for a Hamiltonian formulation. From Eq.~(\ref{pdef}) we have
\begin{equation}
\label{pofu}
\begin{split}
p_\sigma &= \frac{u_\sigma}{z} \bigg[ m + \lambda \omega_f^2 P_{\mu\nu} P^{\mu\nu}
     + \frac{1}{4 \lambda} Q_{\mu\nu} Q^{\mu\nu} \nl
     + \frac{1}{2} E_{\mu\nu} Q^{\mu\nu} \bigg]
     - \bigg[ \delta_\sigma^\alpha + \frac{u_\sigma u^\alpha}{z^2} \bigg]
              C_{\mu\alpha\nu\beta} Q^{\mu\nu} \frac{u^\beta}{z} .
\end{split}
\end{equation}
Using the normalization of the four-velocity $u_\mu u^\mu=-z^2$ leads to the mass-shell constraint
\begin{equation}\label{mshell}
p_{\mu} p^{\mu} + \mathcal{M}^2 = 0,
\end{equation}
with
\begin{align}
\mathcal{M} &= m + H_t , \\
H_t &=
\lambda \omega_f^2 P^{ab} P_{ab}
        + \frac{1}{4 \lambda} Q_{ab} Q^{ab}
        + \frac{1}{2} E^{ab} Q_{ab} , \label{Ht}
\end{align}
where we neglect terms of higher order in curvature and tidal variables.
[Note that this result is analogous to Eq.~(85) in Ref.~\cite{Steinhoff:2014kwa}, but with a factor of 2 typo in the interaction term corrected 
here.] 

The mass-shell constraint (\ref{mshell}) is in fact a special case of a general first-class constraint associated with a gauge freedom. Here, the gauge symmetry is the reparametrization-invariance of the worldline parameter $\sigma$, with its associated gauge freedom represented by the length of $u^\mu$. This leads to the feature that the relation (\ref{pofu}) depends only on the normalized four-vector $u^\mu/z$ and is noninvertible.
Following the usual procedure in constrained dynamics \cite{Dirac:1950pj,Dirac:1964}, the constraint (\ref{mshell})
must be added to the action using a Lagrange multiplier $\alpha$
\begin{equation}\label{SPPcan}
S=\int d\sigma\left[p_{\mu} u^{\mu} + P_{\mu\nu} \frac{D Q^{\mu\nu}}{d \sigma}
 - \frac{\alpha}{2} (p_{\mu} p^{\mu} + \mathcal{M}^2) \right] ,
\end{equation}
with the canonical Hamiltonian being zero. In this form of the action, the function $\alpha$ is undetermined and represents the gauge freedom of the original action (\ref{Raction}). Note also that in (\ref{SPPcan}) the mass-shell constraint takes
the place of the Hamiltonian and all interactions enter as
deformations of the mass shell. This is an important point of view for constructing the test-particle--limit (and then EOB) Hamiltonian, as we shall see in Secs.~\ref{TPL} and \ref{construction}.

\subsection{Imposing the tidal constraints\label{impose}}

In this section we impose the constraints on the tidal variables from the action in Eq.~(\ref{Raction}) through an explicit split into spatial and time components. 

We perform the $3+1$ decomposition of $Q^{\mu\nu}$ and $P_{\mu\nu}$ to single out their spatial SO(3)-irreducible parts. As discussed below Eqs.~(\ref{Qconstraint}) and (\ref{QSTF}), in the body's rest frame the quadrupole is a symmetric-tracefree spatial tensor. Its conjugate momentum $P^{\mu\nu}$ shares the same properties and satisfies the same constraints to linear order in the tidal variables, as can be shown by taking a time derivative of Eq.~(\ref{Qconstraint}) and using the definition (\ref{QEOM}). Therefore, to single out the independent spatial components of $Q^{\mu\nu}$ and $P_{\mu\nu}$ we perform a
Lorentz boost to the rest frame. For this purpose, we project the tidal variables onto a 
tetrad $e_a{}^\mu$, defined such that $g^{\mu\nu} = e_a{}^\mu e_b{}^\nu \eta^{ab}$, where $\eta^{ab}$
is the Minkowski metric. The quadrupole can thus be expressed in terms of its components on the local Lorentz frame $Q^{ab}$ as $Q^{\mu\nu} = e_a{}^\mu e_b{}^\nu Q^{ab}$. The next step is to apply a Lorentz boost $B^{ab}$ to transform
$Q^{ab}$ and $P_{ab}$ to the rest frame which we denote by a tilde,
\begin{align}
  Q_{ab} = B_{ac} B_{bd} \tilde{Q}^{cd} , \qquad
  P_{ab} = B_{ac} B_{bd} \tilde{P}^{cd} .
\end{align}
A particularly simple boost to the rest frame is given by
\begin{equation}\label{standardboost}
B^{ab} = \eta^{ab} - 2 \frac{u^a \delta_0^b}{z}
	+ \frac{(u^a  + z \delta_0^a) (u^b + z \delta_0^b)}{z (z + u^{(0)})} ,
\end{equation}
which is sometimes referred to as a standard boost.
This boost has the properties $B^a{}_b B^{cb} = \eta^{ac}$, $B^a{}_{(0)} = u^a / z$, and $u_a / z B^a{}_b = \eta_{0b}$.
This implies that the constraints
$Q_{\mu\nu} u^\nu = 0$ and $P_{\mu\nu} u^\nu = 0$ become $\tilde{Q}^{a(0)} = 0 = \tilde{P}^{a(0)}$ in the rest frame, where
the round brackets around an index denote the local frame.
The SO(3)-irreducible components of $Q^{\mu\nu}$ and $P_{\mu\nu}$ are therefore the spatial symmetric-tracefree
tensors $\tilde{Q}^{(i)(j)}$ and $\tilde{P}^{(i)(j)}$.
The transformation to the tetrad frame is
\begin{align}
  Q_{ab} &= B_{a(i)} B_{b(j)} \tilde{Q}^{(i)(j)} , \\
  P_{ab} &= B_{a(i)} B_{b(j)} \tilde{P}^{(i)(j)} .
\end{align}

To simplify the notation, we henceforth drop the tilde and the round brackets
for the spatial indices of certain tensors. Specifically, we define
\begin{gather}
Q^{ij} := \tilde{Q}^{(i)(j)} , \quad
P^{ij} := \tilde{P}^{(i)(j)} , \quad
E^{ij} := \tilde{E}^{(i)(j)} , \\
S_Q^{ij} := \tilde{S}_Q^{(i)(j)} = 4 Q^{k[i} P^{j]}{}_k, \label{SQspatial}
\end{gather}
and we also omit the round brackets on indices of $B_{ab}$ since it is
used in the local frame only.

We next consider the split of
the action in Eq.~(\ref{Raction}) into space and time starting with the tidal kinematic term
\begin{equation}
P_{\mu\nu} \frac{D Q^{\mu\nu}}{d \sigma} = P_{ij} \frac{D Q^{ij}}{d \sigma}
+ \frac{1}{2} S_Q^{ij} \frac{D B^{a}{}_i}{d \sigma} B_{aj} .
\end{equation}
Interestingly, the combination of boosts in the last term also
appears in the computation of spin effects \cite{Levi:2015msa}.
Here only the last term of Eq.~(3.18) in Ref.~\cite{Levi:2015msa} gives a
nonvanishing contribution leading to
\begin{align}
\label{boost5}
\frac{D B^a{}_i}{d \sigma} B_{aj} = \frac{u_{(i)} \eta_{ja}}{z + u^{(0)}} \frac{D}{d \sigma} \left[ \frac{u^a}{z} + \delta_0^a \right] - ( i \leftrightarrow j ) .
\end{align}
It is important to note that $\delta_0^j = 0$ can only be inserted in Eq.~(\ref{boost5}) after expanding the
covariant derivative using that
\begin{equation}
\frac{D u_b}{d \sigma} = \dot{u}_b + u^\mu u^a \omega_{\mu a b}.
\end{equation}
Here, the Ricci rotation coefficients are defined to be $\omega_{\mu}{}^{ab}=e^{a\nu}{}_{,\mu}e^{b}{}_{\nu}+e^{b}{}_{\nu}e^{a\rho}\Gamma^{\nu}{}_{\rho\mu}$, the
Christoffel symbols are $2\Gamma_{\alpha\mu\nu}=g_{\alpha\mu,\nu}+g_{\alpha\nu,\mu}-g_{\mu\nu,\alpha}$, and
in this section a dot denotes a derivative with respect to $\sigma$.
Thus, the decomposition of the tidal kinematic terms is explicitly given by
\begin{align}\label{PQdot}
P_{\mu\nu} \frac{D Q^{\mu\nu}}{d \sigma} &= P_{ij} \dot{Q}^{ij} + L_\text{FD} ,
\end{align}
with
\begin{align}
L_\text{FD} &=
u^\mu \omega_{\mu ij} \left[ \frac{S_Q^{ij}}{2} + \frac{S_Q^{ik} u_{(k)} u^{(j)}}{z (z + u^{(0)})} \right] \nnl
- u^\mu \omega_{\mu 0i} S_Q^{ij} \frac{u_{(j)}}{z}
+ \frac{S_Q^{ij} u_{(i)} \dot{u}_{(j)}}{z (z + u^{(0)})} . \label{LFD}
\end{align}
Here FD stands for frame dragging, whose physical origin was explained in the Introduction.
These interaction terms are identical to those in the effective action of ordinary spin effects [as given in Eq.~(5.27) of Ref.~\cite{Levi:2015msa}].

We continue the split into space and time components of the action (\ref{Raction}) with the decomposition of the tidal interaction term,
\begin{equation}\label{LEQ}
L_\text{EQ} = -\frac{z}{2} E_{\mu\nu} Q^{\mu\nu} = - \frac{z}{2} E_{ij} Q^{ij} ,
\end{equation}
where we use the notation $E_{ij} = B^a{}_i B^b{}_j e_a{}^\mu e_b{}^\nu E_{\mu\nu}$.
This term likewise has a corresponding analog in the ordinary spin
calculations given by the spin-induced quadrupole, which is discussed below.
Finally, the oscillator part of the Lagrangian is 
\begin{equation}\label{Lo}
L_o = - z \left[ \lambda \omega_f^2 P_{ij} P^{ij}
 + \frac{1}{4 \lambda} Q_{ij} Q^{ij} \right] .
\end{equation}
Note that the dependence on the metric enters only through the
overall factor $z$, which is the same for the point-mass part, i.e., $- z m$.
Thus, one can view the pure oscillator part as a shift of the mass $m$,
see Appendix \ref{massshift}.

Collecting all the pieces, the matter action is given by
\begin{equation}
\label{S31}
S = \int d\sigma \left[ P_{ij} \dot{Q}^{ij} - m z + L_o + L_\text{EQ} + L_\text{FD} \right] .
\end{equation}
A very similar decomposition can be worked out for the action in Eq.~(\ref{SPPcan}), which is
exercised in Sec.~\ref{TPL}.

\section{Post-Newtonian and test-particle approximations\label{approx}}

In this section we explicitly derive all the terms entering the action (\ref{S31}) 
to 1PN order and obtain the Hamiltonian. The PN approximation requires integrating out the
potential or near-zone ``modes'' of the gravitational field \cite{Goldberger:2004jt, Goldberger:2007hy} 
and usually involves lengthy Feynman integral calculations. Here, however, we can bypass these computations by exploiting 
connections to the point-mass and spin sectors and simply apply certain replacements to
PN results from Ref.~\cite{Levi:2015msa}. Previous results for the action using different approaches were obtained at 1PN order~\cite{Vines:2010ca} and, in the adiabatic limit at 2PN \cite{Bini:2012gu}, see Sec.~\ref{framedrag}.
In addition to the PN limit of the action, we also consider the test-particle limit, which provides information about the strong-field behavior.

\subsection{Potential at 1PN order}

The PN approximation is a weak-field and slow-motion approximation
with orders counted as powers of $v^2 \sim G m/r$, where $v$ is the 
velocity of the object and $r$ is the orbital separation.
In this framework, tidal effects are suppressed by a multipolar approximation parameter, which, 
for the even-parity $2^l$-polar tidal interaction, is given by
\begin{equation}
\label{countmulti}
(\text{$2^l$-pole}) \sim \left( \frac{R}{r} \right)^{2 l + 1} ,
\end{equation}
where $R$ is the object's radius, and $l=2$ for the leading-order tidal effects considered here. For black holes, $R \sim G m$, which means
that the multipolar suppression meshes with the PN power-counting scheme and the conclusion from Eq.~(\ref{countmulti}) is that tidal effects start only at 
5PN order. However, for neutron stars $R\gg Gm$ so that the multipolar scaling fails to mesh with the PN counting. Therefore, tidal effects are considered to start at Newtonian order. In this section, we work out the next-to-leading or first PN (1PN)
corrections to tidal effects and compare to the findings in Ref.~\cite{Vines:2010ca}.

We introduce the following notation. The tidally deformed body is labeled as number 1 and its point-mass companion as
number 2, where the labels are also used for the corresponding masses and orbital
variables. The index $A$ denotes a generic particle label. We continue to give explicit results only for the case of one tidally deformed body, noting that all the expressions can readily be extended to the case of two deformed bodies by adding a copy of all the tidal terms with the particle labels interchanged.
For the worldline parameter, we choose the gauge $\sigma_A = t$, where $t$ is the coordinate time that coincides with the time measured by an asymptotic
observer. 

For the subsequent PN analysis it is convenient to express the action (\ref{S31}) in terms of the PN potential $V_Q$ as
\begin{equation}
\label{SV}
S = \int dt \left[ L_\text{\PM} + P_{ij} \dot{Q}^{ij} - V_Q \right] ,
\end{equation}
where $L_\text{\PM}$ is the point-mass Lagrangian in the PN approximation. Here, the subscript ``pm" denotes a point-particle 
having only a mass monopole. The PN approximated tidal potential is decomposed as 
\begin{equation}
\label{VQ}
V_Q = V_o + V_\text{EQ} + V_\text{FD}, 
\end{equation}
where each part of Eq. (\ref{VQ}) is discussed and derived in detail below. 

The contribution arising from the oscillator (\ref{Lo}) is obtained as follows.  As already noted, the
dependence on the gravitational field enters in this term only through the overall factor of $z$, which is the same
in the point-mass part. We can therefore obtain the PN potential $V_o$ associated with
Eq.~(\ref{Lo}) by a linear shift of the mass $m_1$ in the non-tidal part of the Lagrangian which leads to
\begin{equation}
\label{Vo}
V_o = z_1 \left[ \lambda \omega_f^2 P_{ij} P_{ij}
 + \frac{1}{4 \lambda} Q^{ij} Q^{ij} \right] ,
\end{equation}
where we use
\begin{equation}\label{redshift}
z_A = - \frac{\partial L_\text{\PM}}{\partial m_A} .
\end{equation}
To 1PN order this is explicitly
\begin{equation}
z_1 = 1 - \frac{v_1^2}{2} - \frac{G m_2}{r},
\end{equation}
where $\vct{r} = \vct{y}_1 - \vct{y}_2$, $\vct{v}_A = \dot{\vct{y}}_A$. It is crucial here that $P_{ij}$ is treated as an independent variable, otherwise the dependence of $L_o$ in Eq.~(\ref{Lo}) on the gravitational field would be more complicated. The physical interpretation of the quantity $z_1$ is that it is the 
redshift between the proper time of the worldline $\tau_1$ and the asymptotic observer with time $t$. The fact that the redshift can be obtained from the formula
(\ref{redshift}) was first realized in the context of the first law of mechanics for
binary black holes \cite{Blanchet:2012at}. This can be understood by observing that $L_\text{\PM}$ arises from
the procedure of integrating out the potential modes of the point-mass Lagrangian $- z_A m_A$.
This procedure does not affect the physical meaning of the partial derivative
in Eq.~(\ref{redshift}), hence we have
\begin{equation}
z_A = \sqrt{- u_\mu u^\mu} = \frac{d \tau_A}{d t} .
\end{equation}
In the last step we used the definition of the proper time $d \tau_A^2 = - g_{\mu\nu} dy_A^\mu dy_A^\nu$
and the gauge $\sigma_A = t$. 

Consequently, the PN corrections to the pure oscillator
part have a simple physical interpretation. When the oscillator is
described in terms of the proper time $\tau_1$, it is an ordinary Newtonian
oscillator, in accordance with the expectations from the
equivalence principle. The PN corrections in $V_o$ are due to
the redshift to the asymptotic observer with time $t$, which is used to describe the dynamics in PN
theory and which is measured by the gravitational-wave detectors. This leads to an effective redshift of the resonance frequency away from its value $\omega_f$ measured in the frame of body 1.

The contribution from the interaction terms $V_\text{EQ}$ in Eq. (\ref{VQ}) is associated with the Lagrangian in Eq.~(\ref{LEQ}).
This term is analogous to the spin-induced quadrupole coupling described in Refs.~\cite{Poisson:1997ha, Porto:2008jj, Levi:2015msa}
\begin{equation}\label{LES2}
L_{\text{ES}^2} = \frac{z C_{\text{ES}^2}}{2 m} E_{\mu\nu} S^\mu S^\nu
= \frac{z C_{\text{ES}^2}}{2 m} E_{ij} S^i S^j ,
\end{equation}
where the spin vector is defined by
\begin{equation}
S_{\alpha} = - \frac{1}{2} \eta_{\alpha\beta\mu\nu} \frac{u^{\beta}}{z} \hat{S}^{\mu\nu} , \qquad
S^i = B^{ai} e_a{}^\mu S_\mu ,
\end{equation}
and $\eta_{\alpha\beta\mu\nu}$ is the completely antisymmetric volume form. The spin potentials
are expressed in terms of the canonical spin denoted by a hat. This spin is given by the spatial
components $\hat{S}^i = \frac{1}{2} \epsilon_{ijk} \hat{S}^{(j)(k)}$ of the
spin tensor $\hat{S}^{\mu\nu}$ which satisfies the Newton-Wigner condition
$\hat{S}_{ab}(u^{b}+ z \delta_{0}^{b}) = 0$. Using this condition and the definitions given above, one can
show that $\hat{S}^i = S^i$. Therefore the spins $S^i$ in Eq.~(\ref{LES2}) are those appearing
in the final PN potential. The potential associated with Eq.~(\ref{LEQ}) can
therefore be obtained by substituting the tidal quadrupole in place of the spin quadrupole using the identification
\begin{equation}\label{Qmap}
C_{\text{ES}^2} S^i S^j \rightarrow - m Q^{ij} .
\end{equation}
This further implies the substitution $S^2 \rightarrow 0$ since $Q^{ij}$ is tracefree.
Using these replacements in the 1PN expressions for the spin-induced quadrupole interaction potential given in Eqs. (6.10) and (6.40) of Ref.~\cite{Levi:2015msa} or equivalently in Ref.~\cite{Hergt:2010pa} leads to 
\begin{widetext}
\begin{align}\label{VEQ}
V_\text{EQ} &=
\frac{G m_2}{2r^3} Q^{ij} \big[
         - 3 n^i n^j
	 - v_1^i v_1^j
	 + v_1^i v_2^j
	 + 3 v_1^i n^j ( \vct{v}_1\cdot\vct{n} - \vct{v}_2\cdot\vct{n} )
	 - 3 v_2^i n^j \vct{v}_1\cdot\vct{n}
	 - \tfrac{1}{2} n^i n^j ( 9 v_1^2 
	 - 21 \vct{v}_1\cdot\vct{v}_2
	 + 9 v_2^2 \nl
	 - 15 \vct{v}_1\cdot\vct{n} \vct{v}_2\cdot\vct{n} ) \big]
+ \frac{3}{2} \frac{G^2 m_2}{r^4} ( m_1 + 4 m_2 ) Q^{ij} n^i n^j
+ \frac{G m_2}{r^2} \big[
	 Q^{ij} a_1^i n^j
         + \dot{Q}^{ij} (v_1^i n^j
         - \tfrac{3}{2} v_2^i n^j
	 - \tfrac{3}{4} n^i n^j \vct{v}_2\cdot\vct{n} ) \big] , \nonumber
\end{align}
\end{widetext}
where $\vct{n} = \vct{r} / r$, $\vct{a}_A = \dot{\vct{v}}_A$. This result can readily be extended to 2PN order by applying the spin to tidal-quadrupole mapping (\ref{Qmap}) to the expressions in Ref.~\cite{Levi:2015ixa}.

The last term in Eq.~(\ref{VQ}) describes the interaction of the orbital and tidal angular momentum given in Eq.~(\ref{LFD}), which, as discussed above, is identical to the ordinary spin interaction terms in PN theory.  We can therefore obtain the
corresponding potential $V_{\text{FD}}$ by replacing the spin
by $S_Q^{ij}$ in the PN spin
potentials that are already available. Note that this replacement only works because $S_Q^{ij}$ is independent of the field and instead depends only on the two independent tensors $Q^{ij}$ and $P_{ij}$. Applying the spin to tidal-spin transformation to the leading-order spin-orbit potential
from Eq.~(6.3) in Ref.~\cite{Levi:2015msa} leads to
\begin{equation}\label{VFD}
\begin{split}
V_{\text{FD}} &= -2\frac{Gm_2}{r^2}\vct{S}_Q\cdot\left[\vct{v}_1
\times\vct{n}-\vct{v}_2\times\vct{n}\right] \nl
- \frac{1}{2}\vct{S}_Q\cdot \vct{v}_1\times\vct{a}_1 ,
\end{split}
\end{equation}
where $S_Q^i = \frac{1}{2} \epsilon_{ijk} S_Q^{jk}$. Previous alternative derivations of the result (\ref{VFD}) can be found
in Refs.~\cite{Tulczyjew:1959, Damour:1982, Damour:1988mr}.

This potential is at 1PN order in the tidal case, but in the literature it is usually counted as a 1.5PN spin effect. This happens because the counting of the ordinary spin effects usually assumes a rapidly rotating (extremal) black hole whose spin is considered to be a $0.5$PN contribution, while the tidal spin $\vct{S}_Q$ is fixed by the Newtonian counting of the quadrupole through Eq.~(\ref{SQspatial}).  
The potential could be extended to 3PN order (3.5PN in the ordinary counting) using the results of Refs.~\cite{Hartung:2011te, Levi:2015uxa}.
These interaction terms have the physical interpretation that they describe the frame-dragging 
effect due to a gravito-magnetic field, as explained in Sec.~\ref{framedrag} above.
The intimate connection between spin and frame dragging is evident in the case of a small test spin, which stays constant in a local inertial frame but can change direction
as seen by a distant observer. In general, the particle's worldline deviates from geodesic motion, e.g., due to tidal forces. Thus, the frame associated with the worldline is not inertial, but follows a Fermi-Walker transport. This is encoded in the acceleration-dependent term in Eq.~(\ref{VFD}). Since we consider the case of an irrotational star, we refer to $V_{\text{FD}}$ as frame dragging rather than a spin effect.

The final result for the 1PN tidal Lagrangian (\ref{SV}) is then obtained from Eq.~(\ref{VQ}) together with Eqs.~(\ref{Vo}), (\ref{VEQ}), and (\ref{VFD}). A similar result was derived from the PN equations of motion in Ref.~\cite{Vines:2010ca}. Taking into account the different conventions, we find that the difference between the two expressions is a total time derivative given by
\begin{equation}
\mathcal{L}_Q + V_{\text{EQ}} + V_o
  = \frac{d}{d t} \left[ - \frac{G m_2^2}{r^2 M} Q^{ij} n^i (v_1^j - v_2^j) \right] ,
\end{equation}
where $\mathcal{L}_Q$ is the Lagrangian from Ref.~\cite{Vines:2010ca},
$M = m_1 + m_2$, and we specialize to the adiabatic limit $\omega_f \rightarrow \infty$ in $V_o$.
We can further obtain an equation of motion for the tidal
angular momentum, which at Newtonian order reads
\begin{equation}
\dot{S}_Q^{kl} = \frac{6 G m_2}{r} Q^{j[k} n^{l]} n^j  ,
\end{equation}
in agreement with the tidal torque in Eq.~(1.7) in Ref.~\cite{Vines:2010ca} and our covariant Eq.~(\ref{SQEOM}). While the method based on the PN equations of motion~\cite{Vines:2010ca} and the effective action approach developed here lead to identical results, the advantage of using the effective action is that it makes the underlying structure of the terms (such as the redshift factors) explicit,  and clarifies the relevance of the tidal spin. These insights further facilitate the extension of the results to higher PN orders and the identification of several tidal contributions for which existing results about point-mass and spin potentials can be used.

\subsection{Hamiltonian at 1PN order\label{Hams}}

Implementing dynamical tidal effects in the EOB formalism first requires deriving the Hamiltonian associated with the Lagrangian (\ref{SV}). This can be accomplished by employing a reduction of order to remove higher-order time derivatives
in the potential using the equations of motion \cite{Damour:1990jh}, followed by a Legendre transformation of the velocities.  We apply this procedure to the 1PN tidal Lagrangian in Eq.~(\ref{SV}) using the Newtonian equations of motion
$\dot{Q}^{ij} =2 \lambda \omega_f^2 P_{ij}$. Similarly, to perform the Legendre
transformation it is sufficient to use the Newtonian relations $\vct{v}_A \approx \vct{p}_A / m_A$,
where $\vct{p}_A$ are the canonical momenta conjugate to $\vct{y}_A$.
Since the Hamiltonian can be directly obtained from these substitutions,
we refrain from showing this intermediate result here. Next, we transform to
the center of mass frame where $\vct{p}_1 + \vct{p}_2 = 0$. This results in the following 1PN accurate Hamiltonians
\begin{align}
H_o &= z_1 \left[ \lambda \omega_f^2 P_{ij} P_{ij}
 + \frac{1}{4 \lambda} Q^{ij} Q^{ij} \right] , \label{Ho} \\
z_1 &= 1 - \frac{\vct{p}^2}{2 m_1^2} - \frac{G m_2}{r}, \\
H_{\text{FD}} &= \frac{G}{r^2} \vct{S}_Q \cdot \vct{L}
  \left[ 2 + \frac{3}{2} \frac{m_2}{m_1} \right] , \label{HFD}
\end{align}
\begin{widetext}
\begin{align}\label{H1PN}
H_\text{EQ} &= 
\frac{G}{2m_1 r^3} Q^{ij} \big[
         - 3 m_1 m_2 n^i n^j
	 - \tfrac{3}{2} ( 7 + 3 \tfrac{m_1}{m_2} + 3 \tfrac{m_2}{m_1} ) n^i n^j \vct{p}^2
	 - \tfrac{15}{2} n^i n^j ( \vct{p}\cdot\vct{n} )^2
	 - (1 + \tfrac{m_2}{m_1}) p^i p^j
	 + 3 (2 + \tfrac{m_2}{m_1}) p^i n^j \vct{p} \cdot \vct{n} \big] \nnl
+ \frac{G}{r^2} \lambda \omega_f^2 P_{ij} \big[
	 \tfrac{3}{2} n^i n^j \vct{p}\cdot\vct{n}
         + (3 + \tfrac{2 m_2}{m_1}) p^i n^j \big]
+ \frac{G^2 m_2}{2 r^4} ( 3 m_1 + 10 m_2 ) Q^{ij} n^i n^j,
\end{align}
\end{widetext}
where $\vct{p} = \vct{p}_1 = - \vct{p}_2$ and $\vct{L} = \vct{r} \times \vct{p}$.

To derive the Poisson bracket relations and demonstrate that $Q_{ij}$ and $P^{ij}$ and $ \vct{r}$ and $ \vct{p}$ are canonically conjugate pairs, 
it is useful to consider the action expressed in the form
\begin{equation}\label{Hredaction}
S = \int d\sigma \left[ \vct{p} \cdot \dot{\vct{r}} + P_{ij} \dot{Q}^{ij} - H_\text{\PM} - H_Q \right] ,
\end{equation}
where $H_\text{\PM}$ is the point-mass Hamiltonian in the PN approximation
and the tidal Hamiltonian is
\begin{equation}
H_Q = H_o + H_\text{EQ} + H_\text{FD} .
\end{equation}
The redshift can be obtained from the Hamiltonian through
\begin{equation}\label{redshiftH}
z_A = \frac{\partial H_\text{\PM}}{\partial m_A} .
\end{equation}
Note that the reduction of order in the Lagrangian implicitly also entails a redefinition of the
variables as discussed in Ref.~\cite{Damour:1990jh}. Therefore, the canonical momenta $\vct{p}$
do in general not agree with the spatial components of $p_\mu$.

The equations of motion obtained from varying the action
(\ref{Hredaction}) have the structure of Hamilton's equations and are equivalent to the Poisson brackets
\begin{align}
\{ r^i , p_j \} &= \delta_{ij}, \\
\{ Q^{ij} , P_{kl} \} &= \delta_{ijkl} ,
\end{align}
with all others being zero. The quadrupolar symmetric-tracefree projection operator is given by
\begin{equation}
\delta_{i j k l} = \frac{1}{2} ( \delta_{i k} \delta_{j l}
	+ \delta_{i l} \delta_{j k} ) - \frac{1}{3} \delta_{i j} \delta_{k l} .
\end{equation}
It follows that the tidal angular momentum $S_Q^{ij}$ obeys a canonical SO(3) angular-momentum algebra,
\begin{equation}
\{ S_Q^{ij} , S_Q^{kl} \} = \delta_{ik} S_Q^{jl} - \delta_{jk} S_Q^{il}
- \delta_{il} S_Q^{jk} + \delta_{jl} S_Q^{ik} .
\end{equation}
However, the tidal angular momentum $S_Q^{ij}$ has nonvanishing
brackets with other variables,
\begin{align}
\label{Poisson}
\{ Q^{ij} , S_Q^{kl} \} &= \delta_{il} Q^{kj} - \delta_{ik} Q^{lj}
+ \delta_{jl} Q^{ik} - \delta_{jk} Q^{il} , \\
\{ P_{ij} , S_Q^{kl} \} &= \delta_{il} P_{kj} - \delta_{ik} P_{lj}
+ \delta_{jl} P_{ik} - \delta_{jk} P_{il} .
\end{align}
This algebra implies that $S_Q^{ij}$ is the generator of infinitesimal
rotations for the tidal variables. The interaction terms involving $S_Q^{ij}$ therefore effectively rotate or drag
the frame of the tidal variables. The Poisson brackets (\ref{Poisson}) agree with the bracket algebra for internal degrees of freedom of a spinning
particle in general relativity constructed in Ref.~\cite{Tauber:1987zi}.

\subsection{Test-particle Hamiltonian}
\label{TPL}
The PN approximation discussed above is only valid for slow motion and weak
gravitational fields, but generic mass ratios. By contrast, the
small-mass-ratio approximation is valid for generic velocity and field
strength, but is limited to perturbations of the test-particle limit. The EOB model provides a unified framework to incorporate both the PN and test-particle results in the respective limits, bridging between them. As a first step in building a dynamical tidal EOB Hamiltonian,
in this section, we derive the dynamical tidal Hamiltonian in the test-particle limit following the method in Ref.~\cite{Vines:2016unv}. The adiabatic limit of the tidal Hamiltonian in the test-particle case for circular orbits was computed in Ref.~\cite{Bini:2012gu}.
Furthermore, the frame-dragging contributions can be found in Refs.~\cite{Barausse:2009aa,Vines:2016unv},
where the spin should be mapped to the tidal spin. In order to focus on the new terms in this section, we therefore omit these known frame-dragging
contributions entering via Eq.~(\ref{PQdot}).

We start from the action principle given in Eq.~(\ref{SPPcan}).
Neglecting frame effects in Eq.~(\ref{PQdot}), one can pass to the SO(3)-irreducible
tidal variables by simply replacing 4-indices by local 3-indices. The action
principle then becomes
\begin{equation}
S=\int d\sigma\left[p_{\mu} u^{\mu} + P_{ij} \dot{Q}^{ij}
 - \frac{\alpha}{2} (p_{\mu} p^{\mu} + \mathcal{M}^2)
 + \Order(S_Q) \right] ,
\end{equation}
where as before the dynamical mass is $\mathcal{M} = \mu + H_t$ and the tidal
Hamiltonian is
\begin{align}
\label{Httpl}
H_t =
\lambda \omega_f^2 P_{ij} P_{ij}
        + \frac{1}{4 \lambda} Q^{ij} Q^{ij}
        + \frac{1}{2} E_{ij} Q^{ij} .
\end{align}
We replace the mass $m$ by $\mu$ here for later convenience.
All interactions are encoded in a deformation of the mass-shell constraint,
\begin{equation}\label{mshellTP}
0 = (\mu + H_t)^2 + g^{\mu\nu} p_\mu p_\nu ,
\end{equation}
which follows from the variation of $\alpha$.

We next reduce the orbital variables to their physical components, choosing the coordinate time as the worldline parameter $\sigma = t$,
or $u^0 = 1$, as we did in the PN case. Solving the mass-shell constraint (\ref{mshellTP})
for $p_0$, we obtain the action in the form
\begin{equation}
S = \int dt ( p_i \dot{r}^i + P_{ij} \dot{Q}^{ij} - H_\text{TPL} ),
\end{equation}
where $H_\text{TPL} \equiv - p_0$ is the Hamiltonian in the test-particle limit (TPL). To obtain explicit expressions for the potentials in the Hamiltonian $H_\text{TPL}$ we insert the metric in Schwarzschild coordinates
$(t, r, \theta, \phi)$ into the mass-shell constraint, since similar coordinates are used in the EOB model. The
solution of the mass-shell constraint then leads to
\begin{align}
\label{HTPL1}
H_\text{TPL} &= \sqrt{A_\text{TPL}} \sqrt{(\mu + H_t)^2 + p_e^2 } ,
\end{align}
where $A_\text{TPL} = 1 - 2GM / r$.
We see from Eq.~(\ref{HTPL1}) that the tidal Hamiltonian $H_t$ enters as a shift of the test-mass $\mu$,
which is expected based on the form of the mass-shell constraint (\ref{mshellTP}).
The subscript $e$ on the linear momentum vector denotes that components are
taken in the local tetrad frame $e_a{}^\mu$,
\begin{equation}\label{pe}
\vct{p}_e = \left( p^{(i)} \right)
= \left(
\begin{tabular}{c}
$\sqrt{A_\text{TPL}} p_r$ \\ $p_\theta / r$ \\ $p_\phi / (r \sin \theta)$
\end{tabular}
\right) ,
\end{equation}
and $p_e = |\vct{p}_e|$.
Here the tetrad was chosen as the symmetric matrix square-root of the metric.
Since $H_t$ scales with the fourth power in the mass ratio, we can
expand in $H_t$ as
\begin{align}\label{HTPLexpand}
H_\text{TPL} &\approx \sqrt{A_\text{TPL} ( \mu^2 + p_e^2 ) } + z_\text{TPL} H_t ,
\end{align}
where the redshift is
\begin{equation}\label{zTPL}
z_\text{TPL} = \frac{\partial H_\text{TPL}}{\partial \mu}
\approx \sqrt{A_\text{TPL}} \left[ 1 + \frac{p_e^2}{\mu^2} \right]^{-\frac{1}{2}}
\approx \frac{\mu A_\text{TPL}}{H_\text{TPL}} .
\end{equation}
To derive the expression for the tidal Hamiltonian we decompose it as
\begin{align}
H_t &= H_o^\text{TPL} + H_\text{EQ}^\text{TPL} + \Order(S_Q) , \\
H_o^\text{TPL} &= \lambda \omega_f^2 P_{ij} P_{ij} + \frac{1}{4 \lambda} Q^{ij} Q^{ij} , \\
\begin{split}\label{Htest}
H_\text{EQ}^\text{TPL} &= \frac{3 G M}{2 \mu^2 r^3} Q^{ij} \bigg[
(\vct{n} \times \vct{p}_e)^i (\vct{n} \times \vct{p}_e)^j \nl
- \left( p^{(0)} n^i - \frac{\vct{n} \cdot \vct{p}_e p_{(i)}}{\mu + p^{(0)}} \right)\!
\left( p^{(0)} n^j - \frac{\vct{n} \cdot \vct{p}_e p_{(j)}}{\mu + p^{(0)}} \right)\!
\bigg] ,
\end{split}
\end{align}
where we have obtained $E_{ij} = B^a{}_i B^b{}_j e_a{}^\mu e_b{}^\nu E_{\mu\nu}$ from the mapping in Eq.~(\ref{Qmap})
and Ref.~\cite{Vines:2016unv} in the limit of vanishing Kerr parameter.
In these expressions, one can interchangeably use $u^\mu$ or $p_\mu$ within
our approximation and the quantity $p^{(0)}$ is 
\begin{align}\label{pnperturb}
p^{(0)} = \frac{H_\text{TPL}}{\sqrt{A_\text{TPL}}}
\approx \sqrt{\mu^2 + p_e^2} .
\end{align}
In Eq.~(\ref{Htest})  we have also introduced a vector $(n^i) = (1,0,0)$ so as to express the Hamiltonian in a manifestly rotation-invariant form. This facilitates the use of Cartesian-like frames that are used in PN computations whereas
the frame used above is adapted to spherical coordinates and is therefore noninertial in the Newtonian limit. Hence the frame effects involving $S_Q^{ij}$, which
we ignored in this section, cover not only relativistic frame-dragging
effects, but also Newtonian frame effects (e.g., the Coriolis force).
This becomes more apparent in an explicit calculation below.

Although the test-particle Hamiltonian (\ref{Htest}) is already rather simple, we can make a further
useful approximation, namely that the motion is along circular orbits so that $p_r \approx 0$. In this limit the tidal interaction Hamiltonian reduces to
\begin{equation}\label{HEQTP}
H_\text{EQ}^\text{TPL} = \frac{3 G M}{2 \mu^2 r^3} Q^{ij} \left[
\frac{L^i L^j}{r^2} - \left( p^{(0)} \right)^2 n^i n^j \right] + \Order(p_r) ,
\end{equation}
where $\vct{L} = r \vct{n} \times \vct{p}_e$. An interesting feature of this circular-orbit version of $H_\text{EQ}^\text{TPL}$ is
that no other terms beyond the 1PN approximation appear.

\section{Construction of the effective-one-body Hamiltonian\label{construction}}
In this section we map the above analytical results for relativistic dynamical tides
into the EOB Hamiltonian describing the conservative dynamics of the binary.
The full EOB waveform model, including dissipative effects, is discussed in Ref.~\cite{inprep2}.
The implementation of generic quadrupoles discussed here immediately applies also to spin-induced quadrupoles via Eq.~(\ref{Qmap}), which can be useful for improvements of EOB models for spinning binaries.

\subsection{Structure of the Hamiltonian}
In the EOB approach, incorporating the properties of the bodies other than the masses is non-trivial. For instance, in the case of spinning black holes, different proposals exist~\cite{Taracchini:2013rva,Nagar:2015xqa}. The task of incorporating the effects of dynamical quadrupoles is qualitatively very different from including black-hole spins, since
they further involve the internal dynamics of the bodies. In this section, we therefore elaborate on the basic principles behind the construction of the EOB Hamiltonian to motivate our prescription for incorporating dynamical tidal effects in the EOB model.

The EOB Hamiltonian $H_\text{EOB}$ is based upon an effective Hamiltonian $H_\text{eff}$ describing the motion of an
effective particle in an effective metric \cite{Buonanno:1998gg}. In the test-particle limit, the effective
metric $g^{\alpha\beta}_\text{eff}$ can be chosen as the Schwarzschild or Kerr metric so that the test-particle limit is incorporated in a natural manner. For generic mass ratios, the mapping
between the Hamiltonians is
\begin{equation}\label{HEOB}
H_\text{EOB} = M \sqrt{1 + 2 \nu \left( \frac{H_\text{eff}}{\mu} - 1 \right)} ,
\end{equation}
where $\mu = m_1 m_2 / M$  is the reduced mass and $\nu = \mu / M$.
While alternatives to this map were considered in the literature \cite{Buonanno:1998gg, Damour:2000we},
no compelling reason was found to modify it away from this simple form, except for the one found in
Appendix \ref{massshift} here. The action corresponding to the EOB Hamiltonian is
\begin{equation}\label{SEOB}
S_{\text{EOB}} = \int dt ( p_i \dot{r}^i + P_{ij} \dot{Q}^{ij} - H_\text{EOB} ).
\end{equation}

To construct the effective Hamiltonian it is useful to recall that in the Newtonian limit,
the motion of a binary can be mapped to the motion of a reduced mass $\mu$ in a central
potential of mass $M$. Hence it is natural to start out the EOB construction with a particle of mass $\mu$ moving in a (deformed) effective metric of
mass $M$. The effective Hamiltonian, being a test-particle Hamiltonian,
is then given by $H_\text{eff} = - p_0$ (see Sec.~\ref{TPL}),
where $p_0$ follows as the solution of the mass-shell constraint
\begin{equation}\label{mshelleff}
0 = \mu^2 + \mu^2_\text{NG} + g^{\alpha\beta}_\text{eff} p_\alpha p_\beta .
\end{equation}
Here $\mu_\text{NG}$ incorporates possible effective interactions which lead to a non-geodesic
(NG) motion, analogous to the tidal interactions in Sec.~\ref{TPL}. For the time being, we consider $\mu_\text{NG}$
to be a generic symbol, but assume that a possible dependence on $p_0$ can be treated
perturbatively as in Eq.~(\ref{pnperturb}) when solving the mass-shell constraint.\footnote{Note that $\mu_\text{NG}^2$ is related to the potential $Q$ introduced in Ref.~\cite{Damour:2000we}}
The effective Hamiltonian for a generic effective metric is then given by
\begin{equation}\label{Heff}
H_\text{eff} = - p_0 = \sqrt{A} \sqrt{\mu^2 + \mu^2_\text{NG} + \gamma^{ij}_\text{eff} p_i p_j} + \beta^i p_i ,
\end{equation}
where
\begin{equation}
A = - \frac{1}{g^{00}_\text{eff}} , \qquad
\beta^i = \frac{g^{0i}_\text{eff}}{g^{00}_\text{eff}} ,
\end{equation}
and $\gamma^{ij}_\text{eff}$ is the inverse of the spatial effective metric $g^\text{eff}_{ij}$,
\begin{equation}
\gamma^{ij}_\text{eff} = g^{ij}_\text{eff} - \frac{g^{0i}_\text{eff} g^{0j}_\text{eff}}{g^{00}_\text{eff}} .
\end{equation}
The effective metric and $\mu_\text{NG}$ are fixed by requiring that
$H_\text{EOB}$ agree with the PN and test-particle Hamiltonians in the respective approximations.

\subsection{Matching to the test-particle limit\label{TPEOB}}

Since the foundation for the structure of the effective Hamiltonian is the test-particle limit, we first discuss the inclusion of dynamical tides in the test-particle EOB Hamiltonian. In the test-particle limit the EOB and effective Hamiltonians are related by
\begin{equation}
H_\text{EOB} \approx M + H_\text{eff} - \mu ,
\end{equation}
where the factors of masses are due to the different rest-mass energies of the two Hamiltonians. The test-point-mass limit is
then reproduced by taking the effective metric to be the Schwarzschild metric.

To incorporate the case of a test-particle with dynamical tidal degrees of freedom we consider the mass-shell constraint in Eq.~(\ref{mshellTP}) which is explicitly given by
\begin{equation}
\label{mH}
0 \approx \mu^2 + 2 \mu (H_o^\text{TPL} + H_\text{EQ}^\text{TPL}) + g^{\alpha\beta} p_\alpha p_\beta + \Order(S_Q) ,
\end{equation}
where we have linearized in the tidal terms. Comparing this expression to the constraint given in Eq.~(\ref{mshelleff}) to identify the tidal contributions to the effective metric and the non-geodesic term does not lead to a unique identification. To fix this freedom, we choose the prescription that \emph{all} terms that are quadratic in $p_\mu$ in Eq.~(\ref{mH}) result from a contraction with the effective metric. This implies that $H^\text{TPL}_\text{EQ}$ contributes to the effective metric, which follows from the interaction term in Eq.~(\ref{Httpl}) together with the definition (\ref{defE}) and using that within our approximations $u^\mu$ and $p_\mu$ are interchangeable in this term. The effective metric is then given by
\begin{equation}
g^{\alpha\beta}_\text{eff} = g^{\alpha\beta} + \frac{1}{\mu} C^{\mu\alpha\nu\beta} Q_{\mu\nu} .
\end{equation}
Furthermore, note that the pure oscillator part $H_o^\text{TPL}$ is independent of $p_\mu$ and hence must be included in the non-geodesic term leading to the result $\mu^2_\text{NG} = 2\mu H_o^\text{TPL}$. It is noteworthy that the
effective metric is deformed away from the Schwarzschild metric $g^{\alpha\beta}$ even in the
test-particle limit here, in contrast to non-tidal EOB models where the deformation starts
at linear order in the mass ratio.

Using the above conventions to construct the effective Hamiltonian in Eq.~(\ref{Heff}) 
leads to the following prescription. All the terms in the mass-shell constraint (\ref{mH}) that are quadratic in $p_i$, as seen explicitly by substituting the expression ~(\ref{Htest}) for $H_\text{EQ}^\text{TPL}$, are resummed in $\gamma^{ij}_\text{eff}$, while
all terms linear in $p_i$ and in $p_0$ contribute to the potential $\beta^i$, 
all terms quadratic in $p_0$ contribute to $A$, and all remaining terms are included
in $\mu_\text{NG}$. The importance of having access to additional information from the mass-shell constraint to determine these assignments is highlighted in Appendix \ref{massshift}.

\section{Gauge freedom with dynamical tidal effects\label{gauge}}
Having derived the general structure of tidal contributions to the EOB Hamiltonian based on the test-particle limit, we next discuss several manipulations that are necessary to map the PN results into tidal corrections to the EOB functions. In this section we focus on gauge transformations. We first derive the 1PN accurate general canonical transformation from harmonic to EOB coordinates including the tidal terms. We subsequently apply the method of canonical transformations to obtain a rigorous derivation of the circular-orbit limit. Lastly, we present a convenient choice of frame for the degrees of freedom of the dynamical quadrupole.

\subsection{Tidal terms in the gauge transformations}
To express the PN Hamiltonian $H_\text{PN}$ in the form required by the EOB Hamiltonian
we apply a canonical transformation with generator $g$ to obtain
\begin{equation}\label{gtrafo}
H_\text{EOB} = H_\text{PN} + \{ H_\text{PN}, g \} + \frac{1}{2!} \{ \{ H_\text{PN}, g \}, g \} +{\cal O}(g^3).
\end{equation}
This transformation can be evaluated by making a general ansatz for $g$ and for the PN expansion of the EOB potentials that are invariant under rotations and translations, each involving undetermined coefficients, and solving (\ref{gtrafo}) at each PN order. The solutions for the coefficients are in general not unique, which allows for further simplifications or convenient choices.
The resulting canonical transformation can be viewed as a change of gauge on phase space
with the choices for the free coefficients defining the gauge(s) of the EOB model.

To proceed, we split the canonical transformation into point-mass and tidal parts,
\begin{equation}
g = g_\text{\PM} + g_\text{DT} .
\end{equation}
The point-mass part to 1PN order reads \cite{Buonanno:1998gg}
\begin{align}
\label{gPN}
\begin{split}
g_\text{\PM} &= \frac{\nu r}{2 \mu^2} \vct{p}^2 \vct{n} \cdot \vct{p}
- \frac{G M}{2} \vct{n} \cdot \vct{p} ( 2 + \nu ) .
\end{split}
\end{align}
This generator and the point-mass potentials are uniquely fixed by the requirement
that the effective Hamiltonian is \emph{identical} to the test-particle Hamiltonian,
that is, no corrections in the mass ratio are necessary. At 2PN order, however, this requirement can no longer be satisfied. 
[Yet, the $p_\phi$-dependence in Eq.~(\ref{Heff2}) below can in fact remain unaltered at higher PN orders.
This invariance can be interpreted as a gauge-invariant meaning of the radial coordinate
as the ``centrifugal'' radius \cite{Damour:2014sva}.] 

For the tidal part of the canonical transformation, we choose an ansatz
such that the transformation only generates terms having the same structures as already present in
the 1PN Hamiltonian (\ref{H1PN}). This excludes generators involving
$P_{ij} n^i n^j$ and requires the generator to be linear in $\vct{p}$
and at most quadratic in the tidal variables. This leads to the general form
\begin{equation}
\begin{split}\label{gtide}
g_\text{DT} &= \frac{G m_2}{\mu r^2} Q^{ij} \big[ g_1 n^i n^j \vct{n} \cdot \vct{p} + g_2 n^i p^j \big] \nl
+ \frac{r \vct{n} \cdot \vct{p}}{\mu} \left[ g_3 \lambda \omega_f^2 P_{ij} P_{ij} + \frac{g_4}{4 \lambda} Q^{ij} Q^{ij} \right] \nl
+ \frac{g_5}{\mu^2 r} \vct{n} \cdot \vct{p} \vct{L} \cdot \vct{S}_Q
+ g_\text{circ} ,
\end{split}
\end{equation}
where the coefficients $g_n$ parametrize the freedom in the PN coordinates.
Here, the term involving $g_5$ that is quadratic in $\vct{p}$ is associated with frame effects, which are discussed in detail in Sec.~\ref{corot}
below. To avoid terms of the form $P_{ij} Q^{ij}$, which do not
appear in the Hamiltonian up to 1PN order, we set $g_4 = g_3$. The generator
$g_\text{circ}$ is an additional contribution that is necessary for imposing the circular-orbit limit at the level of the Hamiltonian. It will be discussed in the next section and is given by
\begin{equation}\label{gcirc}
g_\text{circ} = \frac{g_6 G m_2}{\mu r^2 \vct{p}^2} Q^{ij} p^i p^j \vct{n} \cdot \vct{p} .
\end{equation}
Note that this generator should only be used for specializing to circular orbits since it would otherwise produce unusual terms due to the factor of $\vct{p}^2$ in the denominator. 

Another possible term in the generator is the combination
\begin{equation}\label{gosplit}
g_\text{DT} \sim \left[ \frac{\vct{p}^2}{2 \mu} - \frac{GM\mu}{r} \right] P_{ij} Q^{ij} .
\end{equation}
The prefactor here is the Newtonian Hamiltonian, which approximately commutes with $H_\text{PN}$
in Eq.~(\ref{gtrafo}) and therefore does not produce structurally new terms, so that it is formally allowed. However, this transformation
produces terms proportional to the kinetic and potential energies of the oscillator
in Eq.~(\ref{Ho}), but with the opposite relative sign. Since the structure in Eq.~(\ref{Ho})
persists to all PN orders, we can exclude terms like Eq.~(\ref{gosplit}) in the
generator of the canonical transformation. However, for alternative choices of the EOB mapping not considered here, where the kinetic and potential oscillator energies are included in \emph{different}
potentials of the EOB Hamiltonian, the generator in Eq.~(\ref{gosplit}) carries a nonzero coefficient. 

\subsection{Specializing the tidal Hamiltonian to circular orbits \label{qcirc}}
We are ultimately also interested in specializing our results for the tidal terms in the EOB model to circular orbits. In the case considered here, this specialization can be accomplished by starting from the results for generic orbits and substituting $p_r:=\vct{n} \cdot \vct{p}=0$, $\vct{p}^2=L^2/r^2$, and replacing $L^2$ by its value for circular orbits derived from the equations of motion. These ad-hoc substitutions can, however, be justified by employing a rigorous reduction method based on canonical transformations, as detailed below. 

First, we consider the subtleties in the condition for circular orbits. By definition, circular orbits have $r = {\rm const.}$ in time. From the equations of motion for the system (given in Appendix B), it follows that the condition $r={\rm const.}$ concurrently requires the quadrupole degrees of freedom to be in equilibrium. Altogether, this implies that $p_r=0$ for circular orbits in the case considered here.

As mentioned in the previous section, the circular-orbit limit can be imposed through a canonical transformation. It is useful 
to start with general considerations of the effect of the transformation (\ref{gtide}) in the circular limit. Specifically, we note that most of the terms in~(\ref{gtide}) have the structure
\begin{equation}
\label{gf}
g_f = f \frac{r \vct{n} \cdot \vct{p}}{\mu} = f \frac{r p_r}{\mu} ,
\end{equation}
where $f = f(\vct{r}, \vct{p}, Q^{ij}, P_{ij})$ is a generic function of the canonical variables. The effect of a transformation of the form (\ref{gf}) in the circular-orbit limit, to linear order in the tidal variables, and to leading order is
\begin{align}
\{ H_\text{\PM}, g_f \} &= - f \frac{r \dot{p}_r}{\mu} + \Order(p_r), \nonumber \\
&= f \left[ - \frac{\vct{p}^2}{\mu^2} + u \right] + \Order(p_r, u^2), \label{gfresult}
\end{align}
where
\begin{equation}\label{udef}
u = \frac{G M}{r} .
\end{equation}
The transformation (\ref{gfresult}) effectively replaces $\vct{p}^2$ by its value for circular orbits in absence of tidal effects, given by\footnote{The higher-order terms are justified in Appendix~\ref{POLE}.}
\begin{equation}\label{pcirc}
\vct{p}^2 = \mu^2 (u + 3u^2) + \Order(p_r, u^3) .
\end{equation}
To specialize the tidal interaction terms in the Hamiltonian (\ref{H1PN}) to circular orbits one first sets all occurrences of $(\vct{p}\cdot \vct{n})$ to zero. The remaining terms involving $\vct{p}^2$ are eliminated through transformations of the form~(\ref{gfresult}), with $f$ chosen to cancel the coefficient of $\vct{p}^2$ in each case. For example, the circular-orbit limit of the second term in the first line of $H_\text{EQ}$ from Eq.~(\ref{H1PN}) is obtained by using a transformation with $f=-3\mu^2 G Q^{ij}n^i n^j(7+3m_1/m_2+3 m_2/m_1)/(4 m_1 r^3)$. However, generators of the form~(\ref{gf}) are insufficient to remove all the dependences on $\vct{p}$ from the tidal interaction Hamiltonian~(\ref{H1PN}) since there are additional terms having the structure $\sim P_{ij} n^i p^j$ and $\sim Q^{ij}p^i p^j$. To eliminate the former requires a generator of the form $\sim Q^{ij}n^i p^j$ already present in the generic generator (\ref{gtide}) while removing the latter requires a new structure~(\ref{gcirc}) that is absent for generic orbits. For each of these generators the coefficients are chosen so as to remove such terms from the transformed Hamiltonian. Appropriate choices for specific cases are determined in Secs.~\ref{EOBcirc} and \ref{alternatives}.

\subsection{Corotating frame\label{corot}}

In addition to the choice of gauge for the canonical transformations, further freedom remains to choose the frame in which the dynamical quadrupole
components are expressed. This gauge choice on phase space must be treated exactly instead of using
an infinitesimal generator $g$
since it can introduce Newtonian Coriolis forces, and would therefore
require an infinite number of terms in Eq.~(\ref{gtrafo}).
Here, we specialize to a frame that is aligned with the
tidal field in the Newtonian limit. Specifically, this frame is corotating with the orbit
and spanned by the basis vectors $\vct{\Lambda}_I$ given by
\begin{gather}
\label{frameLambda}
\vct{\Lambda}_1 = \vct{n} , \qquad
\vct{\Lambda}_3 = \frac{\vct{L}}{L} = \vct{\ell} , \\
\vct{\Lambda}_2 = \vct{\Lambda}_3 \times \vct{\Lambda}_1
  = \frac{r}{L} (\vct{p} - p_r \vct{n}) .
\end{gather}
We denote the corotating frame by capital indices, as in
\begin{equation}\label{Qrot}
Q^{ij} = \Lambda_I{}^i \Lambda_J{}^j Q^{IJ} , \qquad
P_{ij} = \Lambda_I{}^i \Lambda_J{}^j P_{IJ} .
\end{equation}
The tidal kinematic terms in the EOB action (\ref{SEOB}) then become
\begin{equation}
P_{ij} \dot{Q}^{ij} = P_{IJ} \dot{Q}^{IJ} + S_Q^i \Omega^i ,
\end{equation}
where the angular
velocity of the frame is $\Omega^i = \frac{1}{2} \epsilon_{ikl} \Lambda_I{}^k \dot{\Lambda}_I{}^l$
and reads explicitly
\begin{equation}
\vct{\Omega} = \vct{\Lambda}_1 \times \dot{\vct{\Lambda}}_1 - \vct{\Lambda}_1 \vct{\Lambda}_2 \cdot \dot{\vct{\Lambda}}_3 \label{framerel}.
\end{equation}
The relation (\ref{framerel}) is valid for a generic frame and cyclic
permutations of the frame indices. Substituting the frame (\ref{frameLambda}) leads to
\begin{equation}
\vct{\Omega} = \vct{n} \times \dot{\vct{n}} + \frac{r^2}{L^2} \vct{n} \left(
   \vct{n} \cdot \vct{p} \times \dot{\vct{p}} + p_r \vct{p} \cdot \vct{n} \times \dot{\vct{n}} \right) .
\label{angular}
\end{equation}
We henceforth assume that the tidal quadrupole is aligned with
the orbit and parametrize it as follows
\begin{equation}
\label{eq:corotatingQ}
(Q^{IJ}) = \begin{pmatrix} \alpha+\beta & \gamma & 0\\
        \gamma & \alpha-\beta &  0\\
 0& 0& -2\alpha
       \end{pmatrix} .
\end{equation}
This also implies that the tidal angular momentum is aligned with the
orbital angular momentum, $S_Q^i \sim L^i$, and therefore only the first
term in Eq.~(\ref{angular}) contributes. This term can be eliminated
by a shift of the linear momentum
\begin{equation}\label{pcorot}
\vct{p} \longrightarrow \vct{p} - \frac{1}{r^2} \vct{S}_Q \times \vct{r} ,
\end{equation}
such that
\begin{equation}
p_i \dot{r}^i + P_{ij} \dot{Q}^{ij} \longrightarrow p_i \dot{r}^i + p_\alpha \dot{\alpha}
+ p_\beta \dot{\beta} + p_\gamma \dot{\gamma} .
\end{equation}
Using these results in the EOB action (\ref{SEOB}) implies the new Poisson brackets
for the quadrupole components
\begin{equation}\label{QPB}
1 = \{ \alpha, p_\alpha \} = \{ \beta, p_\beta \} = \{ \gamma, p_\gamma\} ,
\end{equation}
with all others being zero. Here we used the decomposition
\begin{equation}
(P_{IJ}) = \frac{1}{2} \begin{pmatrix}
\frac{p_\alpha}{3} + p_\beta & p_\gamma & 0 \\
p_\gamma & \frac{p_\alpha}{3} - p_\beta & 0 \\
0 & 0 & - \frac{2 p_\alpha}{3}
       \end{pmatrix} .
\end{equation}
This shift produces terms similar to the frame-dragging Hamiltonian (\ref{HFD}),
but depends on $\vct{p}$. For this reason it is useful to include the term involving $g_5$
in Eq.~(\ref{gtide}).

We find it most convenient here to first map the PN results to
the EOB potentials and then transform the EOB action to the corotating frame.
The effect of the rotation from Eq.~(\ref{Qrot}) in the tidal terms of the EOB Hamiltonian
can be obtained from the relations
\begin{equation}
\vct{p} = p_r \vct{\Lambda}_1 + \frac{L}{r} \vct{\Lambda}_2 , \qquad
\vct{n} = \vct{\Lambda}_1 ,
\end{equation}
and the orthonormality of the basis $\vct{\Lambda}_I$.
Within our approximations, the transformation in Eq.~(\ref{pcorot}) is only applied to the point-mass terms, i.e., to the linear momentum terms under the
square root in Eq.~(\ref{Heff2}) since $\vct{S}_Q$ is already quadratic in the
dynamical tidal variables. The effect of the transformation in Eq.~(\ref{pcorot})
can then be written as a contribution to $\mu$ of the form
$\mu_\text{frame}^2 = - 2r^{-2} \vct{S}_Q \cdot \vct{L} + \Order(S_Q^2) $.
However, since this contribution is linear in $\vct{p}$ it should be
rewritten as a contribution of the form $f_\text{frame}=\beta^i_\text{frame} p_i / \mu$ as
\begin{equation}\label{fframe}
f_\text{frame} = - \frac{A_\text{\PM}}{H_\text{eff,\PM} \mu r^2} \vct{S}_Q \cdot \vct{L}
   + \Order(S_Q^2).
\end{equation}
Note that, aside from the linearization in $S_Q$, this equation is exact.

\section{Effective-one-body Hamiltonian for dynamical tidal effects\label{DTEOB}}
In this section, we use the results of the previous sections to derive the EOB model
for dynamical tidal effects. We first devise the model for generic orbits before discussing the specialization to circular orbits. In the case of point masses this reduction introduces poles at the light-ring into the Hamiltonian. We discuss the use of gauge transformations to understand the origin of such poles and options for their removal, and further show that the tidal model developed here is free of such pathologies. We also explore several alternative prescriptions for including the tidal information in the EOB model to demonstrate that the importance of dynamical tides is not an artifact of the
particular choice of the EOB resummation of tidal effects.

\subsection{Generic orbits at 1PN order\label{EOBgen}}
Before proceeding with the presentation of our tidal EOB Hamiltonian, we introduce convenient notations
for the ingredients of the effective Hamiltonian (\ref{Heff}).
We split the potential $A$ into point mass (``pm") and dynamical-tidal parts
\begin{align}
A = A_\text{\PM} + A_\text{DT}.
\end{align}
For the potential $\beta^i$ the point-mass terms
vanish by our assumption that both bodies are nonspinning, however there is a contribution from the tidal frame effects given by
\begin{equation}
f_\text{DT} = \frac{\beta^i p_i}{\mu} .
\end{equation}
For the tidal terms in the other EOB functions $\mu_\text{NG}$ and $\gamma^{ij}_\text{eff}$ we use the fact that the tides are a small correction to the point-mass case and collect all the dynamical tidal terms into a single function $\mu_\text{DT}$ given by
\begin{equation}
\mu_\text{NG}^2 + \gamma^{ij}_\text{eff} p_i p_j = \mu^2_\text{DT} + \mu_\text{\PM}^2 + \frac{p_r^2}{D_\text{\PM}}
+ \frac{L^2}{r^2} .
\end{equation}
Here $p_r = \vct{n} \cdot \vct{p}$ and $p_\phi = L$, which agree with the Schwarzschild momenta for $\theta = \pi / 2$
and $p_\theta = 0$. The point-mass parts of the potentials can be found
in Eq.~(2) of  Ref.~\cite{Taracchini:2013rva} and Eq.~(10) of Ref.~\cite{Taracchini:2012ig}, and are summarized in Appendix \ref{PMEOB}.
With these conventions our ansatz for the dynamical tidal extension of
the EOB Hamiltonian is
\begin{align}\label{Heff2}
H_\text{eff} &= \sqrt{A \left[ \mu^2 + \mu^2_\text{DT} + \mu^2_\text{\PM} + \frac{L^2}{r^2}  +\frac{A p_{r}^2}{D_\text{\PM}}\right] }\nnl
+ \mu f_\text{DT} .
\end{align}
The quantities $A_\text{DT}$, $\mu_\text{DT}$, and $f_\text{DT}$ are determined below by matching to the
PN Hamiltonian up to a canonical transformation, and they are independent of
the linear momentum.

To construct the tidal EOB potentials $A_\text{DT}$, $\mu_\text{DT}$,
and $f_\text{DT}$ we express them as
\begin{align}
A_\text{DT} &= \mathcal{E}_{ij} Q^{ij}, \qquad
f_\text{DT} = - Z \vct{S}_Q \cdot \vct{\ell} , \label{ADT} \\
\frac{\mu^2_\text{DT}}{\mu^2} &= \frac{z_c}{2\mu \lambda}\left(Q^{ij}Q^{ij}+4\lambda^2 \omega_f^2 P_{ij}P_{ij}\right)+Q^{ij}{\cal C}_{ij}.\label{muDT}
\end{align}
The quantities $\mathcal{E}_{ij}$, $\mathcal{C}_{ij}$, $Z$, and $z_c$ are defined below. We do not include interaction terms involving $P_{ij}$,
although they appear in Eq.~(\ref{H1PN}), i.e., we assume that $P_{ij}$ only appears
in the oscillator kinetic energy in the EOB Hamiltonian. This condition, which restricts the
gauge freedom, is suggested not only by the test-particle limit in Eq.~(\ref{HEQTP}), but also by the structure of the covariant coupling in Eq.~(\ref{LQ}), where the tidal field $E^{\mu\nu}$ couples only to $Q^{\mu\nu}$, but not to $P_{\mu\nu}$.
In fact,  the terms involving $P_{ij}$ in Eq.~(\ref{LQ}) arise from partial integrations
in the PN computation and could be avoided by making different choices of the residual gauge freedom.

We next posit an ansatz for $\mathcal{E}_{ij}$, $\mathcal{C}_{ij}$, $Z$, and $z_c$
that is fixed by requiring that the PN expansion of $H_\text{EOB}$ agrees with the Hamiltonians
from Sec.~\ref{Hams} up to a canonical transformation.
The canonical transformation is required since, in general, the PN Hamiltonians do not fit into the EOB structure. As the last step, we transform the EOB Hamiltonian to the
corotating frame, that is, we add Eq.~(\ref{fframe}) to $f_\text{DT}$.
The tidal interaction is then encoded in
\begin{align}
\label{ECgeneric}
\mathcal{E}_{ij} &= -\frac{3G m_2}{\mu r^3} n^i n^j\left\{ 1 - [2 X_2 - (1-c_1) \nu] u \right\} , \\
\mathcal{C}_{ij} &= \frac{3 G m_2}{\mu^3 r^3} \bigg\{ \frac{L^2}{r^2} \ell^i \ell^j
                   + \left[ 1 + (c_2-2c_1) \nu \right] n^i p^j p_r \nnl 
                   + \left[ (1-c_1) \vct{p}^2 + (5c_1-c_2) p_r^2 \right] \nu n^i n^j \bigg\} , \label{Ctide}
\end{align}
the correction to the redshift factor is
\begin{equation}
\begin{split}
z_c &= 1+\frac{3}{2} X_1 u 
    + \frac{\nu}{2} (1+2c_1) \left[ \frac{\vct{p}^2}{\mu^2} - u \right] ,
\end{split}
\end{equation}
and the frame effects are described by
\begin{equation}
\begin{split}\label{Zspin}
Z &= \frac{L}{\mu^2 r^2} \bigg\{ 1+\left[3X_1-5-(1+c_2)\nu\right]\frac{u}{2} \nl
     - (1 - c_2 \nu) \frac{\vct{p}^2}{2 \mu^2} - c_2 \nu \frac{p_r^2}{\mu^2} \bigg\} ,
\end{split}
\end{equation}
where $X_A = m_A / M$. The remaining gauge freedom is contained
in the arbitrary constants $c_1$ and $c_2$. The gauge parameters are explicitly given by
\begin{subequations}\label{ggeneric}
\begin{align}
g_1 &= \frac{3 \nu}{4} (2c_1-1) , &
g_2 &= 1 + \frac{X_1}{2} , \\
g_3 &= g_4 = \frac{X_1}{2} + (1-2c_1) \frac{\nu}{2} , &
g_5 &= \frac{c_2 \nu}{2} ,
\end{align}
\end{subequations}
and $g_6 = 0$.

Recall that we work in the corotating frame, so that contractions
of the quadrupole should be replaced by the canonical variables
$\alpha$, $\beta$, and $\gamma$, with the Poisson brackets given in Eq.~(\ref{QPB}) and using the relations
\begin{align}
Q^{ij} n^i n^j &= \alpha + \beta , &
Q^{ij} n^i p_j &= p_r (\alpha+\beta) + \frac{L}{r} \gamma , \label{Qnnrel} \\
Q^{ij} \ell^i \ell^j &= - 2 \alpha , &
\vct{S}_Q \cdot \vct{\ell} &= 2 (\beta p_\gamma - \gamma p_\beta) ,
\end{align}
\begin{align}
P_{ij} P_{ij} &= \frac{p_\alpha^2}{6} + \frac{p_\beta^2}{2} + \frac{p_\gamma^2}{2} , \\
Q^{ij} Q^{ij} &= 6 \alpha^2 + 2 \beta^2 + 2 \gamma^2 . \label{QQrel}
\end{align}

\subsection{Circular orbits and 2PN completion\label{EOBcirc}}

In this section, we specialize the dynamical tidal EOB model to circular orbits using a canonical transformation to remove occurrences of the linear momentum from the Hamiltonian, as described in Sec.~\ref{qcirc}. We also discuss the inclusion of information at 2PN
order in the model.

Following the method in Sec.~\ref{qcirc}, the tidal
terms (\ref{ECgeneric})--(\ref{Zspin}) for circular orbits simplify to be
\begin{subequations}
\label{allcirc}
\begin{align}
\mathcal{E}_{ij} &= -\frac{3 G m_2}{\mu r^3} n^i n^j\left\{ 1 - [2 X_2 - (1-c_1) \nu] u + \mathcal{E}_\text{2PN} u^2 \right\} , \label{Ecirc} \\
\mathcal{C}_{ij} &= \frac{3 G^2 m_2}{\nu r^4} ( 1+ 3u ) \left[ \ell^i \ell^j
                   + (1-c_1) \nu n^i n^j \right], \label{Ccirc} \\
\begin{split}
z_c &= 1+\frac{3}{2} X_1 u \left[ 1 + \frac{9}{4} u \right] , \label{zcirc}
\end{split}\\
\begin{split}
Z &= \frac{L}{\mu^2 r^2} \bigg\{ 1 + \left[3X_1-6-\nu\right]\frac{u}{2} \nl
   - [ X_1 (9+6\nu) + \nu (3+\nu)] \frac{u^2}{8} \bigg\} . \label{Zcirc}
\end{split}
\end{align}
\end{subequations}
The gauge parameters used to obtain these expressions are given by
\begin{align}
g_1 &= \frac{3}{4} (\nu-2) , &
g_2 &= 1 + \frac{X_1}{2} , \\
g_3 &= g_4 = \frac{X_1}{2} + \nu , &
g_5 &= \frac{1}{2} ,
\end{align}
with a nonvanishing coefficient in Eq.~(\ref{gcirc}) equal to
\begin{equation}
g_6 = -\frac{3}{2} .
\end{equation}
The fact that a nonvanishing generator $g_\text{circ}$ is required to eliminate the momenta from the Hamiltonian implies that taking
the circular-orbit limit and performing the EOB resummation do not commute here,
since Eq.~(\ref{gcirc}) is not admitted as a canonical transformation
for generic orbits. 
We note that the remaining free parameter $c_1$ in Eq.~(\ref{allcirc}) is not related to a gauge parameter $g_n$
here, but can be used to move a term between $\mathcal{E}_{ij}$ and $\mathcal{C}_{ij}$.
It is chosen such that the result in this section follows from that of the
previous section by inserting the circular-orbit expression for $\vct{p}^2$ given in Eq.~(\ref{pcirc}).
However, as discussed in the previous section, such an insertion is in general not a correct procedure, in contrast to
the adapted canonical transformation involving the term in Eq.~(\ref{gcirc}).

The expressions (\ref{allcirc}) already include information at 2PN order determined in the following way. The 2PN terms in the redshift, Eq.~(\ref{zcirc}), follow from Eq.~(\ref{redshiftHEOB}), while the 2PN correction in
Eq.~(\ref{Zcirc}) is a combination of the spin-orbit frame-dragging terms in Ref.~\cite{Damour:2008qf}
and the corotating frame addition in Eq.~(\ref{fframe}). For the tidal interaction terms we have added a parameter
$\mathcal{E}_\text{2PN}$ to Eq.~(\ref{Ecirc}). In general, one would expect such 2PN corrections
also in $\mathcal{C}_{ij}$, but for simplicity we do not consider this modification; the 2PN terms in
Eq.~(\ref{Ccirc}) arise only from substituting the linear momentum for circular orbits from
Eq.~(\ref{pcirc}). We fix $\mathcal{E}_\text{2PN}$ by using the results for adiabatic tidal (AT) effects in the EOB model that were calculated to 2PN order in Ref.~\cite{Bini:2012gu}. In that model, 
all adiabatic quadrupolar tidal effects are included in the potential $A$, by setting $A=A_\text{\PM}+A_\text{AT}^\text{2PN}$ with
\begin{multline}
A_\text{AT}^\text{2PN} = -\frac{3\lambda X_2 G^2 M}{X_1 r^6} \bigg[ 1 + \frac{5}{2}X_1 u \\
      + \left(\frac{337}{28}X_1^2+\frac{1}{8}X_1+3\right)u^2 \bigg] . \label{AAT2PN}
\end{multline}
Requiring that the adiabatic limit of our model discussed in Appendix \ref{equilibrium} gives the same result for the 2PN expansion of $H_{\rm EOB}$ as that obtained from using Eq.~(\ref{AAT2PN}) determines that
\begin{equation}
\mathcal{E}_\text{2PN} = \frac{5 X_1}{28} (33X_1-7) .
\end{equation}
 Note that while by construction the PN expansion of our model agrees with the PN expansion of the results of Ref.~\cite{Bini:2012gu} a nonperturbative specialization of our EOB model to adiabatic tides does not reproduce the EOB model in
Ref.~\cite{Bini:2012gu}, which we explain in Sec.~\ref{EOBLR}.

We further note that it is not possible to completely remove the linear momentum from all terms
using a canonical transformation. In particular, the frame term (\ref{Zcirc}) is still linear in
$L$. Inserting the circular-orbit relation
\begin{equation}\label{Lcirc}
\frac{L}{\mu r} = \sqrt{u} \left[ 1 + \frac{3 u}{2}
+ \left( \frac{27}{4} - 3\nu \right) \frac{u^2}{2} \right] + \Order(p_r) ,
\end{equation}
in $Z$ leads to
\begin{equation}
\begin{split}\label{Zcirc2}
Z &= \frac{u^{3/2}}{G M\mu} \bigg\{ 1 - (3 X_{2}+\nu)\frac{u}{2} \nl
   - [X_2(9-6\nu)+\nu(27+\nu)] \frac{u^2}{8} \bigg\} .
\end{split}
\end{equation}
In general, substituting relations like Eq.~(\ref{Lcirc})
in the Hamiltonian is not justified, since they are derived using the equations of motion.
But, as long as the tidal spin is small $S_Q \approx 0$, as is the case in the adiabatic
limit, inserting Eq.~(\ref{Lcirc}) in the Hamiltonian amounts to adding an approximate
``double zero'' to the Hamiltonian, which is legitimate \cite{Barker:1980:1,Barker:1980:2}.
This means that while inserting Eq.~(\ref{Lcirc}) alters the equation of motion
for the orbital phase, the change is proportional to $S_Q$ and hence negligible, provided that the assumption that $S_Q$ is small is valid.
Nevertheless, it is important to keep terms linear in $S_Q$ in the Hamiltonian, since
they also influence the equations of motion for the dynamical quadrupole in the
form of frame effects, as discussed in Sec.~\ref{framedrag}.

From the discussion above, it is obvious that a specialization to circular orbits relies on several
assumptions. Furthermore, the used circular-orbit relations are 2PN exact only and are not exact in the test-particle limit. However, when considering the final 25 cycles of the inspiral waveform for several binary configurations we find that the difference between using the circular- and generic-orbit tidal terms is small compared to the uncertainty due to the lack of higher-order PN information.

\subsection{Behavior near the light ring\label{EOBLR}}

In the test-particle limit, the light ring is the
(marginally stable) circular orbit for a massless particle such as a photon and is
located at $u=1/3$ in the Schwarzschild spacetime. Its importance for test-particle motion is that. when specializing the Hamiltonian to circular orbits. most quantities exhibit a pole at this location due to the value of $\vct{p}^2$ being
\begin{equation}\label{circularTPL}
\vct{p}^2 = \frac{\mu^2 u}{1 - 3 u} + \Order(p_r).
\end{equation} 
%


Previous, adiabatic tidal EOB models~\cite{Bini:2012gu,Bernuzzi:2014owa} that incorporated
test-particle and gravitational self-force results specialized the tidal potentials to
circular orbits and thus introduced poles into the Hamiltonian. In Ref.~\cite{Bernuzzi:2014owa} the pole marks the location of an approximate\footnote{The light ring is determined in Ref.~\cite{Bernuzzi:2014owa} from an approximate EOB model
that only incorporates PN tidal results since a self-consistent solution for
the light ring of the tidal EOB model of Ref.~\cite{Bernuzzi:2014owa} is difficult to obtain~\cite{Bini:2012gu}.}
light ring that is shifted away from the test-particle--limit value due to corrections coming
both from the mass ratio and PN tidal interactions.
These singularities are problematic in an EOB evolution especially for neutron-star--black-hole
binaries with large mass ratios, where the orbit may pass through the
pole before the end of the inspiral, which leads to an unphysical divergence. 
  
As originally pointed out in Sec.~VII.C of Ref.~\cite{Akcay:2012ea} and explained in detail in our Appendix~\ref{POLE}, the pole in the tidal contributions to the Hamiltonian is due to a pathological choice of gauge, but as we discuss in Appendix~\ref{POLE}
it can be eliminated through a canonical transformation.
The gauge choice made in Refs.~\cite{Bini:2012gu,Bernuzzi:2014owa} that
gave rise to the pole is a consequence of the requirement that tidal terms are independent of
the linear momentum for circular orbits, or equivalently
that $r$ is the ``centrifugal'' radius \cite{Damour:2014sva}. This means that in this gauge the function $L^2=\vct{p}^2r^2+{\cal O}(p_r)$ appears in the effective Hamiltonian only as the combination $A L^2/r^2$,
like in the point-mass Hamiltonian in the Schwarzschild background. In the model developed here, this gauge choice is unavailable
due to the richer structure of the couplings involving $L^2$ for a generic quadrupole such as the $\ell^i \ell^j$-term in
Eq.~(\ref{Ctide}). This term is invariant under the residual gauge
freedom parametrized by $c_1$ and $c_2$ and therefore cannot be removed to reproduce the gauge of Ref.~\cite{Bini:2012gu}. Note that our tidal EOB model (\ref{ECgeneric}) reproduces the test-particle limit case from Eq.~(\ref{HEQTP}) without introducing any explicit singularities.

\subsection{Alternative factorized resummations\label{alternatives}}

To account for the uncertainty due to lack of complete knowledge of the dynamical tidal effects beyond 1PN
order, we develop different prescriptions for incorporating PN tidal information in the EOB Hamiltonian. In particular, we consider two alternatives where all corrections are included either in $A_\text{DT}$ or in $\mu_\text{DT}^2$. For each case, we devise both a factorized form and a Taylor expanded version. Comparing the gravitational waveforms generated based on these different EOB Hamiltonians allows us to assign an uncertainty to our model.

We start by considering the case where all tidal corrections are included in $\mu_\text{DT}^2$.
Mimicking the structure of the covariant interaction terms in Eq.~(\ref{RQ}) we express $\mu_\text{DT}^2$ in the form
\begin{equation}\label{mufact}
\frac{\mu^2_\text{DT}}{\mu^2} = \frac{2 z_{\mu^2}}{\mu} \bigg[
\frac{Q^{ij}Q^{ij}}{4\lambda} + \lambda \omega_f^2 P_{ij}P_{ij}
+ \frac{1}{2}Q^{ij}E_{ij}
- g_Q \vct{S}_Q \cdot \vct{\ell} \bigg] ,
\end{equation}
with $A_\text{DT} = 0 = f_\text{DT}$.
The reparametrization-invariance of Eq.~(\ref{RQ}) requires an overall
factor of $z$, which corresponds to $z_{\mu^2}$ here. For the tidal field $E^{ij}$,
we assume that it is given by the test-particle expression with an
overall factor $E_c$ accounting for the finite mass-ratio PN corrections, such that
\begin{equation}
E_c H_\text{EQ}^\text{TPL} =\frac{1}{2} Q^{ij} E_{ij} ,
\end{equation}
or explicitly
\begin{align}\label{Efact}
E_{ij} &= \frac{3 G M}{ \mu^2 r^3} E_c \left[ \frac{L^2}{r^2} \ell^i \ell^j
                   - ( \mu^2 + \vct{p}^2 ) n^i n^j
      + p_r n^i p_j \right] .
\end{align}
This agrees with Eq.~(\ref{HEQTP}) for $E_c = 1$, and the term involving $p_r$ reproduces the 1PN
expansion of  Eq.~(\ref{Htest}). Note that within our approximations it would also be consistent to 
use the test-particle results Eq.~(\ref{Htest}) in Eq. (\ref{Efact}) instead of its 1PN expansion, which could potentially lead to further improvements of the model, but is not considered here.

Following the same procedure as before, namely requiring that the PN expansion of $H_\text{EOB}$ --- using Eqs.~(\ref{mufact}), (\ref{Efact}), and $A_\text{DT} = 0 = f_\text{DT}$ --- agrees with the PN Hamiltonian from Sec.~\ref{Hams} up to a canonical transformation, we determine the factors in the Hamiltonian to be
\begin{align}
E_c &= \frac{\mu}{m_1} \left\{ 1 + \frac{X_1}{2} u + E_\text{2PN} u^2
      - \frac{3 \nu}{2} \left[ \frac{\vct{p}^2}{\mu^2} - u - 3 u^2 \right] \right\} , \\
\begin{split}\label{gQ}
g_Q &= \frac{L}{\mu r^2} \bigg\{ 1 - (3 + \nu) \frac{u}{2}
  - (9 + 9 \nu + \nu^2) \frac{u^2}{8} \nl
  - \frac{\nu}{2} \left[ \frac{\vct{p}^2}{\mu^2} - u - 3 u^2 \right] \bigg\} ,
\end{split}\\ \label{zmu}
z_{\mu^2} &= 1 + \frac{3 X_1}{2} u + \frac{27 X_1}{8} u^2
            + \frac{\nu}{2} \left[ \frac{\vct{p}^2}{\mu^2} - u - 3 u^2 \right] .
\end{align}
The gauge parameters read
\begin{subequations}\label{gfactor}
\begin{align}
g_1 &= - \frac{3 \nu}{4} , &
g_2 &= 1+ \frac{X_1}{2} , \\
g_3 &= g_4 = \frac{X_1}{2} + \frac{\nu}{2} , &
g_5 &= g_6 = 0.
\end{align}
\end{subequations}
The 2PN completion is
\begin{equation}
E_\text{2PN} = \frac{36}{7} X_1^2 - \frac{13}{8} X_1 ,
\end{equation}
determined by matching the PN expansion of the adiabatic result in Ref.~\cite{Bini:2012gu} for circular orbits.

We refer to the model in the form (\ref{mufact}) as a factorized model due to the overall
factor of $z_{\mu^2}$ and the factor of $E_c$ in the tidal field.
Interestingly, the factorized structure leaves no free gauge parameters.
Hence it can be considered as a gauge independent representation at 1PN order. It would be interesting to include a similar factorization into our EOB model from Sec.~\ref{EOBgen} with the aim of singling out a unique
gauge. We leave this for future work, where we will also compute
the 2PN dynamical tidal effects.

Next, we consider an EOB model where all tidal terms are included in the
potential $A$ as in Ref.~\cite{Bini:2012gu}. This is very similar to Eq. (\ref{mufact}), with the modification that the overall factor $z_A$ is different. Specifically,
\begin{equation}\label{Afact}
A_\text{DT} = \frac{2 z_A}{\mu} \bigg[
\frac{Q^{ij}Q^{ij}}{4\lambda} + \lambda \omega_f^2 P_{ij}P_{ij}
+ \frac{1}{2} Q^{ij}E_{ij}
- g_s \vct{S}_Q \cdot \vct{\ell} \bigg] ,
\end{equation}
with $\mu^2_\text{DT} = 0 = f_\text{DT}$ and
\begin{align}\label{zAexact}
z_A &= \frac{\mu^2 A^2}{H_\text{eff}^2} z_m \\
\begin{split}\label{zA}
&\approx 1 + \frac{3}{2} u (X_1 - 2) - \frac{9 X_1}{8} u^2 \nl
+ \left[ \frac{\nu}{2} - 1 \right] \left[ \frac{\vct{p}^2}{\mu^2} - u - 3 u^2 \right] .
\end{split}
\end{align}

It is important to emphasize that, although above we have included terms at 2PN order, for the case of generic orbits the expressions should be truncated at
1PN order. This is because the 2PN terms were matched to the results of Ref.~\cite{Bini:2012gu}, and hence are only consistent for circular orbits. The specialization of the above results for $A_\text{DT}$ and $\mu^2_\text{DT}$ to circular orbits is accomplished following the procedure described in Sec.~\ref{EOBcirc}. This amounts to inserting Eqs.~(\ref{pcirc}) and (\ref{Lcirc}) while keeping the factorized structure.
The model from Eq.~(\ref{mufact}) reproduces the test-particle limit by construction, while the model from Eq.~(\ref{Afact}) achieves this only when the exact expression
for the redshift correction $z_A$ (\ref{zAexact}) is used.

Finally, for both EOB models presented in this section, we also consider a
Taylor expanded version, where the entire expressions are expanded and
the result is truncated to the desired PN order (1PN or 2PN). In the adiabatic limit,
the 2PN Taylor expanded version of $A_\text{DT}$ reduces to the model in Ref.~\cite{Bini:2012gu} given in Eq.~(\ref{AAT2PN}).

\subsection{Effective Love number for dynamical tides\label{keff}}

Adding the degrees of freedom for a dynamical
quadrupole to the EOB model considerably increases the computational cost
to generate waveforms. In this section, we develop a model that
eliminates the quadrupole variables while still capturing the effect
of dynamic tides. This is achieved through the effective tidal
deformability function introduced in Sec.~\ref{tideintro} and derived
below. Due to its computational advantages, this model is currently
being implemented as the TEOB model in the LIGO Algorithm Library (LAL) used
for searches and parameter-estimation studies.

The effective model for the dynamical tides is based on approximate
solutions for the quadrupole degrees of freedom for a Newtonian
inspiral. To obtain these solutions we perform a systematic
multi-timescale analysis as
described in the textbook~\cite{kevorkian2012multiple}. However, since
these computations are not particularly illuminating and merely follow
standard methods we refrain from giving the details here. Instead, we
present a simplified analysis that exhibits the main features and
results.

For the subsequent calculations it is more convenient to parametrize the quadrupole in the body frame instead of in the corotating frame as  
\be
Q_{ij}=\begin{pmatrix} \alpha+b & c & 0\\
        c & \alpha-b &  0\\
 0& 0& -2\alpha
       \end{pmatrix},
\ee
and to transform to the variables in the corotating frame after obtaining the solutions. The relation between the variables in the two frames is
\be
\beta=\cos(2 \phi) b+\sin(2\phi)c, \ \ \gamma=-\sin(2\phi)b+\cos(2\phi)c . \ \ \ \ \label{eq:beta}
\ee
Using the body-frame variables, the Newtonian conservative equations of motion for circular orbits are given by
\be
\label{eq:eomb}
\begin{Bmatrix}\ddot \alpha\\
\ddot{b}\\
 \ddot c\end{Bmatrix}
+\omega_f^2 \begin{Bmatrix}\alpha\\
b\\
             c  \end{Bmatrix}
 = \omega_f^2{\cal A}(r) \begin{Bmatrix}1/3\\
\cos (2\phi)\\
 \sin (2\phi)\end{Bmatrix}, \ \ \ \ \ \ \ \ \  
\ee
where $\phi=\int \Omega(t) dt$ is the orbital phase and
\begin{equation}\label{Acal}
 {\cal A}(r(t)) = \frac{3 Gm_2 \lambda}{2 r^3} ,
\end{equation}
is the amplitude of the tidal force. The dynamical degrees of freedom $b$ and $c$, calculated using the method of variation of parameters and  trigonometric identities, are given by
\bea
\frac{2}{\omega_f}\begin{Bmatrix} 
 b\\
c
\end{Bmatrix}
&=& \cos(\omega_f t)\int dt {\cal A}\begin{Bmatrix}  \sin(2\phi+\omega_f t)\\
\cos(2\phi+\omega_f t)\end{Bmatrix}\nonumber\\
&+& \sin(\omega_f t)\int dt {\cal A} \begin{Bmatrix} \cos(2\phi+\omega_f t)\\  \sin(2\phi+\omega_f t) \end{Bmatrix} \nonumber\\
&+&\cos(\omega_f t)\int dt {\cal A} \begin{Bmatrix}  \sin(2\phi-\omega_f t)\\
- \cos(2\phi-\omega_f t)\end{Bmatrix}\nonumber\\
&+& \sin(\omega_f t)\int dt {\cal A} \begin{Bmatrix} \cos(2\phi-\omega_f t)\\  \sin(2\phi-\omega_f t) \end{Bmatrix} \nonumber\\\
&+&\begin{Bmatrix} a_1^b\\ a_1^c \end{Bmatrix} \cos(\omega_f t)+\begin{Bmatrix} a_2^b\\ a_2^c \end{Bmatrix}  \sin(\omega_f t),
 \label{eq:decomp} \ \ \ \ \ \ \ \ \ \ 
\eea
where the terms in the last line are homogeneous solutions. 
The functions $\Omega(t)$ and $r(t)=(GM)^{1/3}\Omega(t)^{-2/3}$ in the Newtonian approximation evolve on the radiation-reaction timescale, which we assume to be slow compared to the orbital timescale. Therefore, locally in time ${\cal A}$ can be treated as constant. In this limit, the solution for the static component $\alpha$ is $\alpha={\cal A}/{3}$. For the dynamical degrees of freedom, a resonance between the tidal forcing and the f-modes occurs when $\Omega\sim \omega_f/2$ or $(2\phi-\omega_f t)\sim 0$. In the regime away from the resonance and assuming that $r$ evolves slowly, the integrals (\ref{eq:decomp}) can be performed as they stand, and the solutions for no initial mode excitation reduce to
\be
\begin{Bmatrix}b^{\rm outer}\\ c^{\rm outer}\end{Bmatrix}=\frac{{\cal A}}{1-\frac{4\Omega^2}{\omega_f^2}}\begin{Bmatrix} \cos(2\phi)\\
                                                   \sin(2\phi)
                                                  \end{Bmatrix}
. \ \ \ \ \label{eq:outersol}
\ee
Transforming to the corotating frame, the outer solution for $\beta$ obtained from Eqs.~(\ref{eq:beta}) and (\ref{eq:outersol}) is
\be
\beta^{\rm outer}=\frac{{\cal A} }{1-\frac{4\Omega^2}{\omega_f^2}}. \label{eq:betaouter}
\ee
Near the resonance, however, the last two integrands in Eq.~(\ref{eq:decomp}) have a stationary phase and require a specialized treatment such as a Taylor expansion around the resonance which takes into account the evolution of $\Omega$ due to radiation reaction. We define a small parameter
\be
\label{epsilon}
\epsilon=\frac{64}{5}2^{1/3}G^{2/3}M^{2/3}\omega_f^{5/3}\mu ,
\ee
that characterizes the ratio of the radiation-reaction and orbital timescales at the resonance. The frequency has the near-resonance expansion
\be
\Omega\approx \frac{\omega_f}{2}+\dot \Omega(t_f) (t-t_f)+ {\cal O}\left((t-t_f)^2\right).
\ee
Here, $t_f$ is computed by integrating $\dot r=-64\mu G^2M^2/(5 r^3)$ up to the resonance radius $r_f^3=4GM/\omega_f^2$. 
The phase in the integrands in Eq.~(\ref{eq:decomp}) is then
\be
\chi\equiv 2\phi-\omega_f t \approx \chi_f+ \dot \Omega (t-t_f)^2+{\cal O}\left((t-t_f)^3\right). \label{eq:expandchi}
\ee
The integrands are stationary as long as $\chi-\chi_f$ is small. When $\chi-\chi_f={\cal O}(1)$ they are again oscillatory, indicating that the system has left the resonance's region of influence. The duration of the resonance can thus be estimated from 
\be
1\sim (\chi-\chi_f)\sim \dot\Omega (t-t_f)^2,
\ee
which implies that the resonance lasts for a time $t_{\rm res}\sim 1/\sqrt{\epsilon}$ since $\dot \Omega={\cal O}(\epsilon)$. 
To conveniently describe the near-resonance behavior, we use the phase instead of time as the dependent variable and introduce a rescaled variable 
\be
\hat t =\sqrt{\epsilon}(\phi-\phi_f) =\frac{8 \left(1-\frac{r^{5/2}\omega_f^{5/3}}{2\ 2^{2/3} G^{5/6}M^{5/6}}\right)}{5
   \sqrt{\epsilon }}\label{eq:thatdef}.
\ee
Using the expansion in Eq.~(\ref{eq:expandchi}), the definition from Eq.~(\ref{eq:thatdef}), and transforming to the corotating frame using ~(\ref{eq:beta}) leads to the near-resonance solution
\bea
\beta^{\rm res}
&=& \frac{\tilde {\cal A}}{\sqrt{\epsilon}} \bigg[\cos(\Omega^\prime \hat t^2)\int_{-\infty}^{\hat t} \sin(\Omega^\prime s^2)ds\nonumber\\
&& \ \ \ \ \ \ \ -\sin(\Omega^\prime \hat t^2)\int_{-\infty}^{\hat t} \cos(\Omega^\prime s^2)ds\bigg], \ \ \ \ \ \ \ \ \ \ \  \label{eq:betares}
\eea
where $\tilde{\cal A}={\cal A} \omega_f^2 / (4 \Omega^2)=3 m_2 \omega_f^2\lambda/(8 M)$, $\Omega^\prime=3/8$ is a rescaled derivative of $\Omega$, and the factor of $1/\sqrt{\epsilon}$ arises from converting $dt$ to $d\hat{t}$. The lower limit of the integrals refers to times long before the resonance. To construct a composite solution that incorporates both the resonance and the outer behavior involves adding the two solutions and subtracting their common term to avoid double-counting. This common term can be identified by expanding Eq.~(\ref{eq:betares}) for $\hat t\to -\infty$ and expanding the outer solution~(\ref{eq:betaouter}) for $\Omega \to \omega_f/2$, taking into account the slow evolution of $\Omega$ and using the definition~(\ref{eq:thatdef}). The results are
\be\lim_{\Omega\to \omega_f/2}\beta^{\rm outer}=-\frac{\tilde {\cal A}}{2\sqrt{\epsilon}\hat t \Omega^\prime}=\lim_{\hat t\to -\infty}\beta^{\rm res}.\ee
The two solutions match and the composite solution is
\begin{widetext}
\be\label{betaDT}
\beta^{\rm DT}=\frac{\tilde{\cal A} }{\frac{\omega_f^2}{4\Omega^2}-1}+\frac{\tilde{\cal A}}{2\sqrt{\epsilon}\hat t \Omega^\prime}+\frac{\tilde {\cal A}}{\sqrt{\epsilon}} \bigg[\cos(\Omega^\prime \hat t^2)\int_{-\infty}^{\hat t} \sin(\Omega^\prime s^2)ds-\sin(\Omega^\prime \hat t^2)\int_{-\infty}^{\hat t} \cos(\Omega^\prime s^2)ds\bigg].
\ee
\end{widetext}
Note that we consider here only the behavior up to frequencies of $\omega_f+ {\cal O}(\sqrt{\epsilon})$ which fails to describe the dynamics long after the resonance, but is sufficient for the range of frequencies reached during a binary inspiral. 

Using the solution~(\ref{betaDT}) we compute $\lambda_{\rm eff}$ defined in Eq.~(\ref{leffdef}) as the ratio to the adiabatic result. 
The adiabatic solution for $\beta$ is obtained by expanding Eq.~(\ref{eq:betaouter}) for $4\Omega^2/\omega_f^2\to 0$ and gives $\beta^{\rm AT}={\cal A}$. This leads to
\be\label{keffeq}
\frac{\lambda_{\rm eff}}{\lambda}=\frac{1}{4}+\frac{3}{4}\frac{\beta^{\rm DT}}{{\cal A}}.
\ee
The expression~(\ref{keffeq}) can be converted to a function of  the orbital radius $r$ and the tidal parameters by using in the result for $\beta^{\rm DT}$ from Eq.~(\ref{betaDT}) the definitions~(\ref{eq:thatdef}) and~(\ref{epsilon}), together with the relation $\Omega^2=GM/r^3$ and the value $\Omega^\prime=3/8$. The integrals in Eq.~(\ref{betaDT}) are standard Fresnel integrals (e.g., they are available in Mathematica with the convention $\int_{-\infty}^{\hat t }\sin(\Omega^\prime s^2)ds=\frac{\sqrt{\pi}}{2\sqrt{2\Omega^\prime}}[1+2\, {\rm FresnelS}(\hat t\sqrt{2\Omega^\prime}/\sqrt{\pi})]$.)

To incorporate this result in the EOB model we first consider the connection to the adiabatic limit more generally. 
From the quadrupole equation of motion given by
$\ddot{Q}^{ij} + \omega_f^2 Q^{ij} = - \lambda \omega_f^2 E_{ij}$ one can verify 
the identity
\begin{equation}
\frac{\lambda_\text{eff}}{4} E_{ij} E_{ij} =
- \frac{Q^{ij}}{4 \lambda \omega_f^2} \left[ \ddot{Q}^{ij}
  + \omega_f^2 Q^{ij} + 2 \lambda \omega_f^2 E_{ij} \right] .
\end{equation}
Here, the left-hand side is the coupling used to obtain the 2PN
adiabatic tidal interaction in Ref.~\cite{Bini:2012gu}, while the
right-hand side is identical to the tidal Lagrangian from
Eq.~(\ref{LNewton}) except for the first term, which differs by an
irrelevant total time derivative. This implies that we can obtain a
dynamical tidal EOB model by starting with the adiabatic EOB model
from Ref.~\cite{Bini:2012gu} given in Eq.~(\ref{AAT2PN}) here and replacing $\lambda \rightarrow
\lambda_\text{eff}$ using Eqs.~(\ref{k2}) and ~(\ref{keffeq}).

\begin{table}
\begin{tabular}{ll}
Model & Equations \\
\hline
TEOB-$A_\text{AT}$ & (\ref{AAT2PN}) \\
TEOB-$k_\text{eff}$ & (\ref{AAT2PN}) with $\lambda_\text{eff}$, (\ref{keffeq}), (\ref{betaDT}), (\ref{Acal}) \\
TEOB & (\ref{ADT}), (\ref{muDT}), (\ref{Ecirc})--(\ref{zcirc}), (\ref{Zcirc2}) \\
TEOB-$\mu_\text{DT}^f$ & (\ref{mufact}), (\ref{Efact})--(\ref{zmu}) \\
TEOB-$A_\text{DT}^f$ & (\ref{Afact}), (\ref{zA}), (\ref{Efact})--(\ref{gQ})
\end{tabular}
\caption{We list the tidal EOB models that we consider in Figs.~\ref{dphi} and \ref{dphi2}, 
together with the equations that define them (see for all cases also the energy map (\ref{HEOB}) and 
the effective Hamiltonian (\ref{Heff2})). The formulas should be specialized to circular orbits by inserting
Eqs.~(\ref{pcirc}) and (\ref{Lcirc}), if applicable. An explicit form in terms
of canonical quadrupole variables in the corotating frame with Poisson brackets
from Eq.~(\ref{QPB}) is obtained through the relations in
Eqs.~(\ref{Qnnrel})--(\ref{QQrel}). The superscript ``f'' stands for ``factorized.''
We also consider models where the factorization is expanded and PN truncated. Those models are 
denoted by a superscript ``e.''\label{modtab}}
\end{table}
\begin{figure*}
\begin{minipage}{0.5\linewidth}
\centering \includegraphics[width=\linewidth]{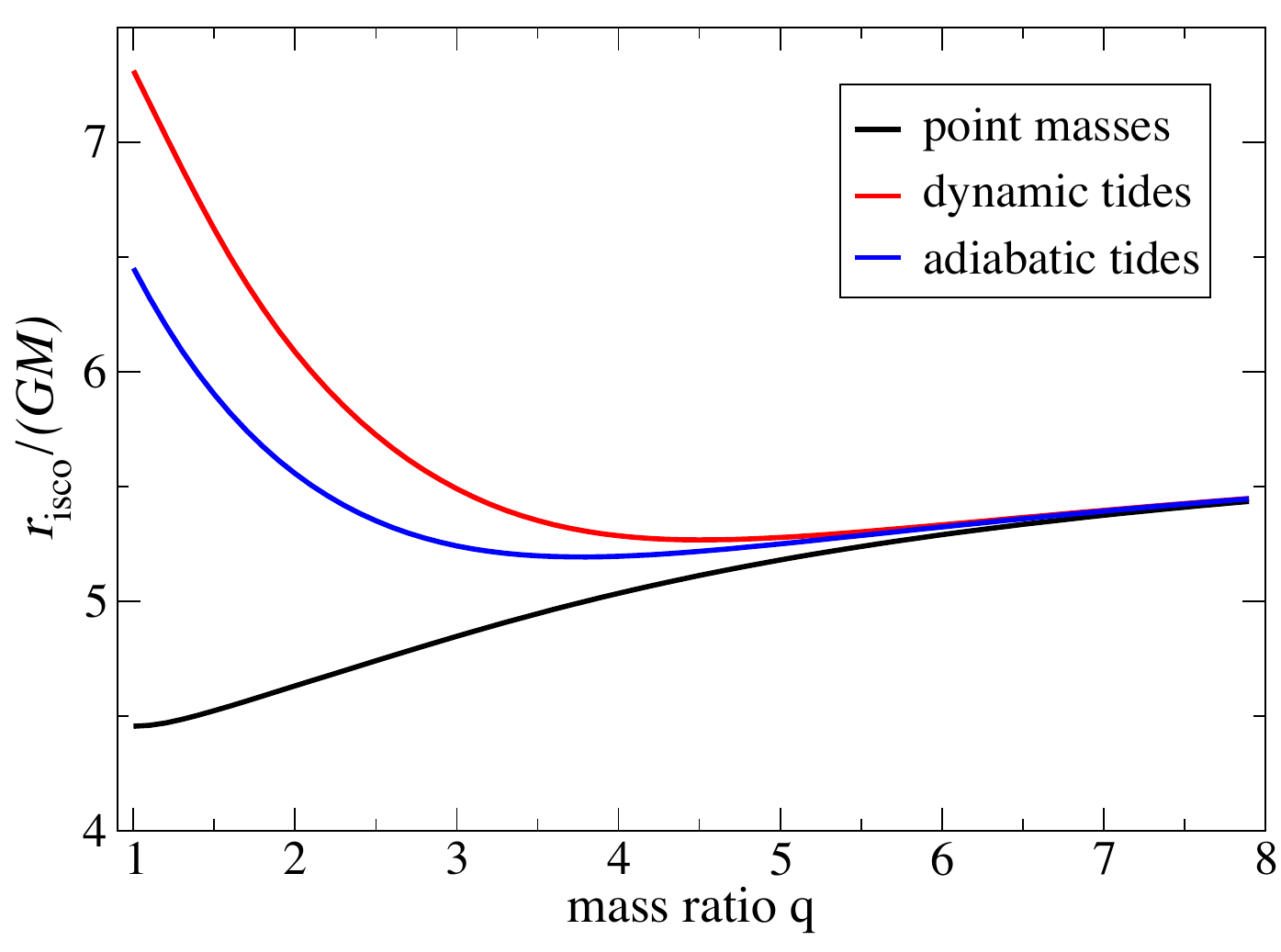}
\end{minipage}%
\begin{minipage}{0.5\linewidth}
\centering \includegraphics[width=\linewidth]{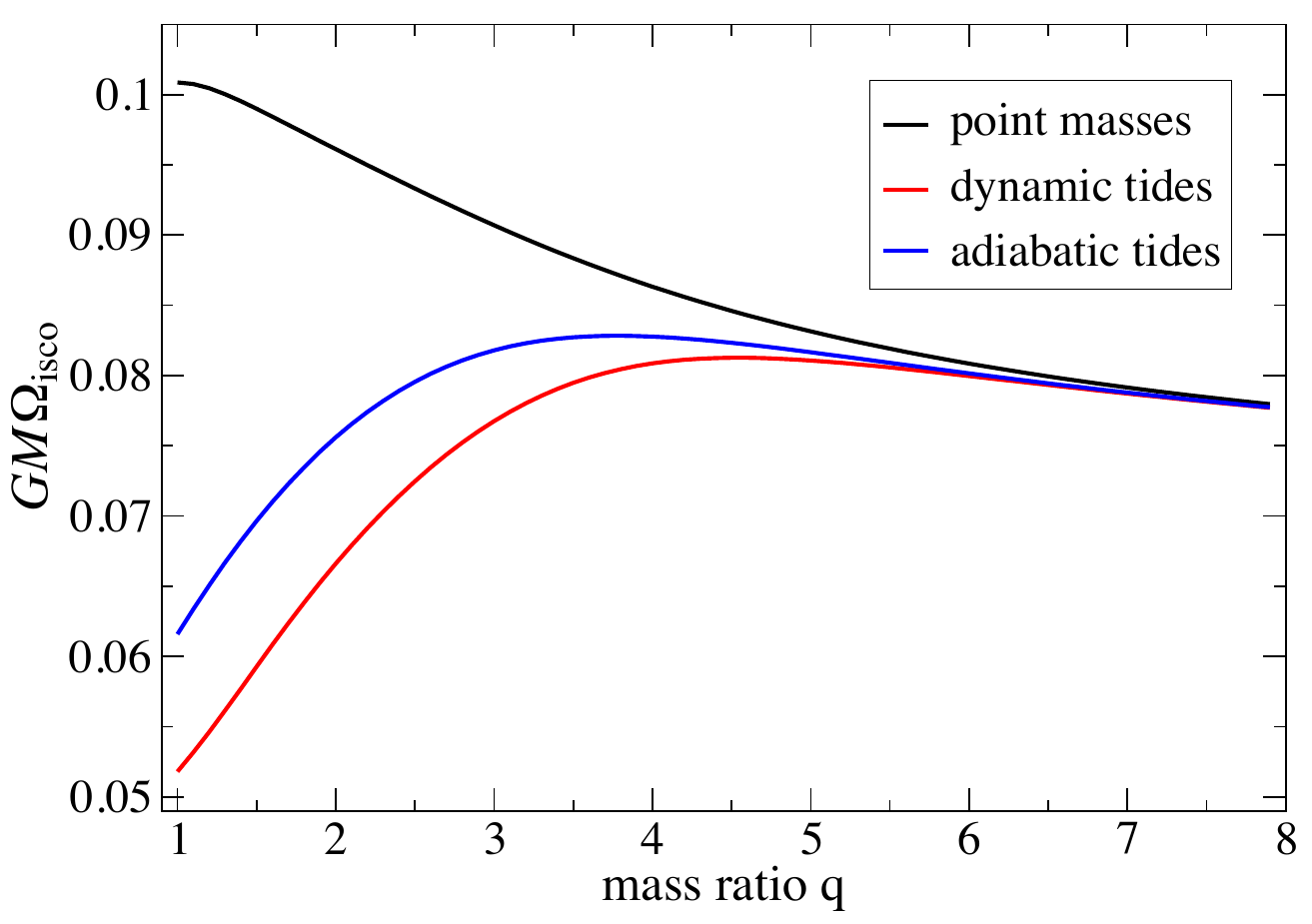}
\end{minipage}
\caption{Innermost stable circular orbit (ISCO) as a function of the mass ratio for a neutron-star--black-hole binary. As soon as the adiabatic tidal effects deviate
from the point-mass case, the dynamical tidal effects are relevant, too. Here we used the
2PN accurate TEOB-$k_\text{eff}$ model with an effective Love number from Sec.\ \ref{keff}.\label{isco}}
\end{figure*}
\begin{figure*}
\begin{minipage}{0.5\linewidth}
\centering \includegraphics[width=\linewidth]{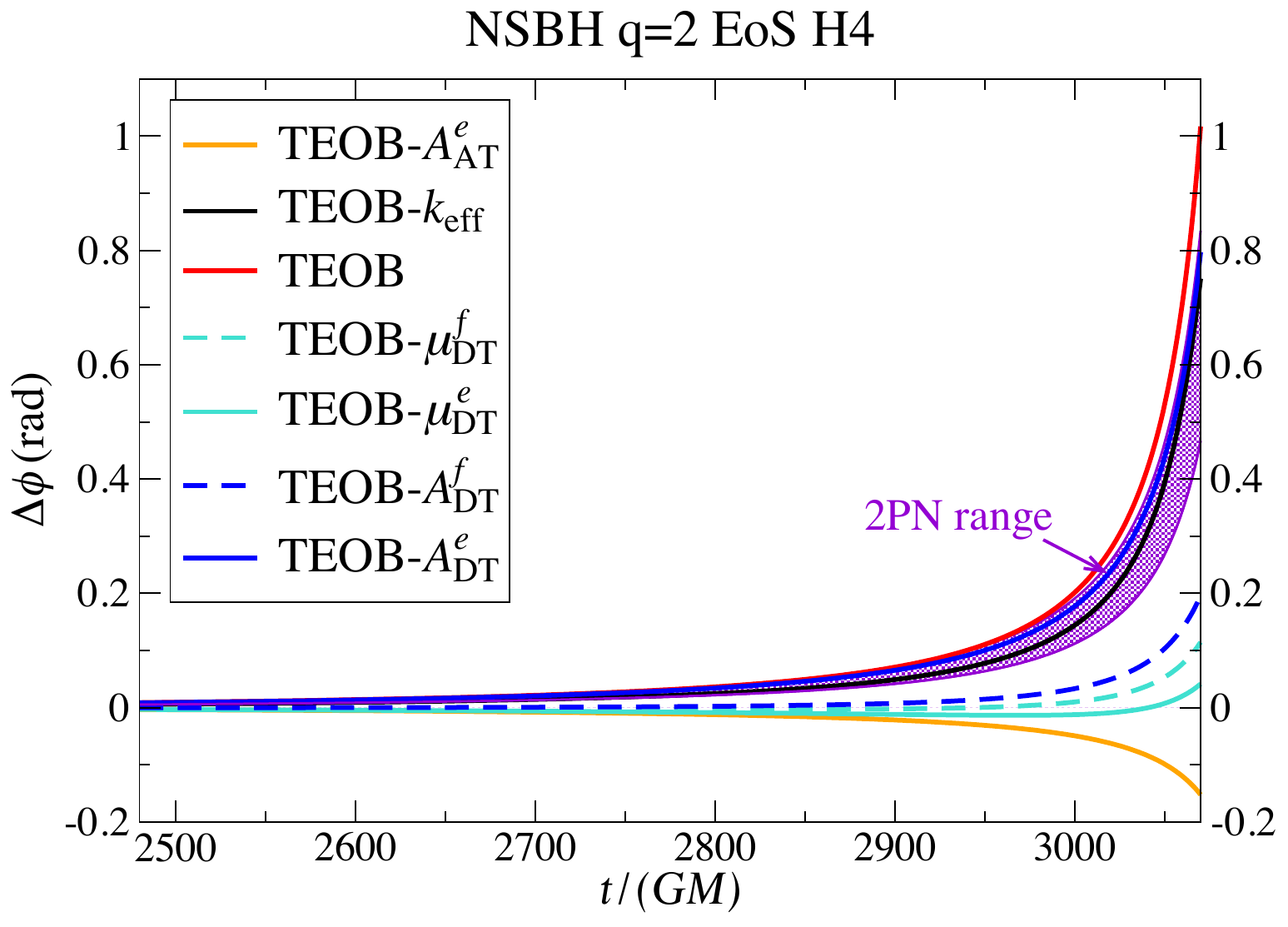}
\end{minipage}%
\begin{minipage}{0.5\linewidth}
\centering \includegraphics[width=\linewidth]{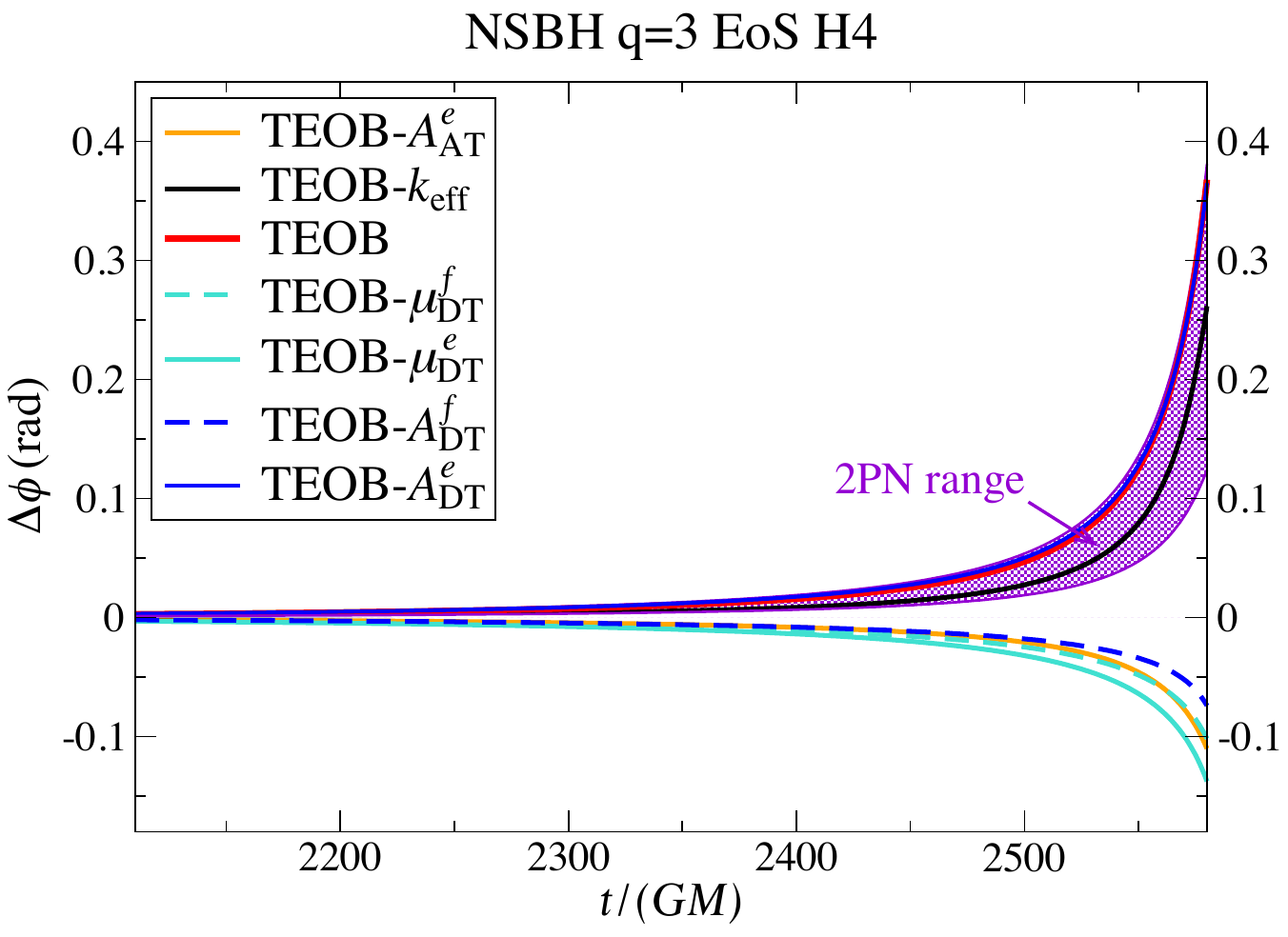}
\end{minipage}
\caption{Phase difference in radians between waveforms using the 2PN TEOB-$A_\text{AT}$ model \cite{Bini:2012gu}
as the baseline and the models summarized in Table \ref{modtab} for $m_1=1.350 M_\odot$ 
and a piecewise polytropic approximation of the H4 equation of state. While individual lines are shown
for the 1PN truncation of the models, the shaded area  encompasses the range of all dynamical models at 2PN order.
The fact that the span with 2PN information lies within the 1PN span indicates that
our conclusions about the importance of dynamical tides will likely remain valid when higher PN orders are included. Furthermore, the TEOB model (red curve) is always close to the upper part of the span.
\label{dphi}}
\end{figure*}
\begin{figure}
\centering \includegraphics[width=\linewidth]{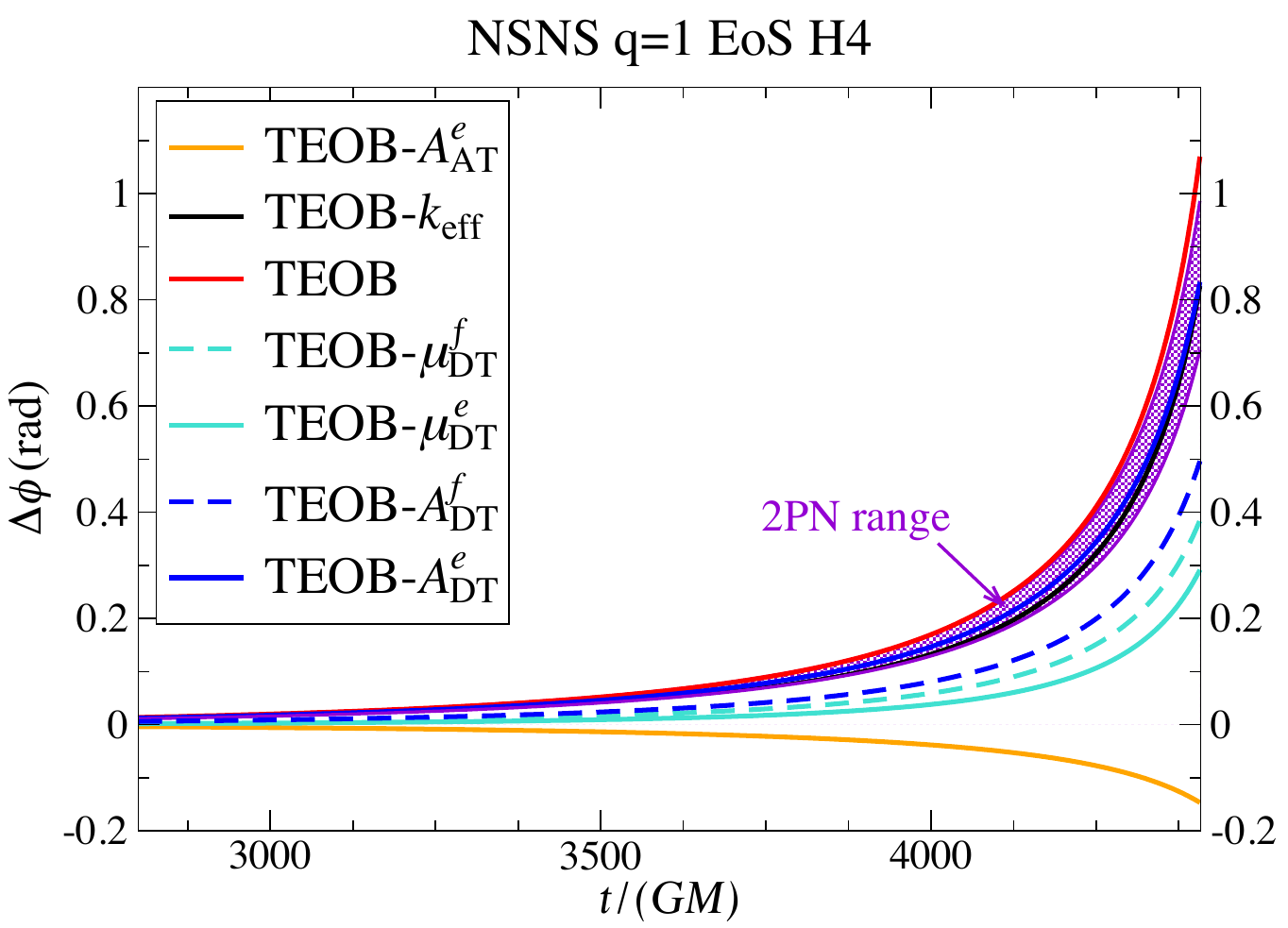}
\caption{Same as Fig.\ \ref{dphi}, but for an equal-mass neutron-star binary.
\label{dphi2}}
\end{figure}

\section{Results and Discussion}
\label{discuss}

Before assessing the importance of dynamical tidal effects, we give more details
about the EOB model used in the analysis. We only consider the circular-orbit version of all
results here, since in this case more information about tidal effects at 2PN order is available. 
We checked that the generic-orbit version of the Hamiltonian typically
differs by less than 0.1 radian from the circular-orbit version (for circular orbits at 1PN order and over 24 cycles).
We fix the remaining arbitrary constant in the model in Eq.~(\ref{allcirc}) to be
 $c_1 = 0$, since this choice implies that the gauge parameters~(\ref{ggeneric}) are the same as for the factorized models~(\ref{gfactor}).
This choice can be revised once the complete 2PN dynamical result becomes available.
The initial conditions for the EOB evolutions are the equilibrium solutions of the EOB
equations of motion determined numerically, as explained in Appendix \ref{equilibrium}.

We first consider the effect of dynamical tides in the conservative dynamics, specifically on the location of the innermost
stable circular orbit (ISCO). This is shown in Fig.~\ref{isco} for the case $m_1 = 1.350 M_\odot$ with a radius of 13.5~km and a piecewise polytropic approximation to the H4 equation of state
from Refs.~\cite{Read:2008iy,Hotokezaka:2013mm}, which gives a tidal deformability of $\lambda G / ( G m_1)^5 = 1111$ and f-mode 
frequency of $ G m_1 \omega_f = 0.0629$. The plots suggest that dynamical tidal effects become important as soon as tidal effects become 
relevant. 

However, the most interesting observables are the gravitational waves emitted
by the system. To generate waveforms, we also include radiation-reaction forces constructed  from the flux in
Ref.~\cite{Taracchini:2012ig} together with the analytically known adiabatic tidal corrections to the 
waveform modes from Ref.~\cite{Damour:2012yf}, as explained in detail in Ref.~\cite{inprep2}. We evaluate the relevance of dynamical tidal effects on gravitational waves using
the following models devised in the previous section:
(i) the dynamical tidal model based on the EOB construction developed in Sec.~\ref{EOBgen} and denoted by TEOB here,
(ii) the factorized models from Sec.~\ref{alternatives} where tidal terms are contained exclusively in either $\mu_\text{DT}^2$ or $A_\text{DT}$ and denoted by TEOB-$\mu_{\rm DT}^{f}/A_{\rm DT}^{f}$, (iii) their Taylor expanded versions TEOB-$\mu_{\rm DT}^{e}/A_{\rm DT}^{e}$, and (iv) the 
``effective--Love-number'' model denoted by TEOB-$k_{\rm eff}$ based on Sec.~\ref{keff} and the adiabatic EOB
model (\ref{AAT2PN}) from Ref.~\cite{Bini:2012gu}. These models are summarized in Table~\ref{modtab}.

The difference in the gravitational-wave phase between our dynamical tidal models and
the 2PN TEOB-$A_\text{AT}$ adiabatic tidal model from Ref.~\cite{Bini:2012gu}
used as a baseline is shown in Figs.~\ref{dphi}
and \ref{dphi2}. Note that the span of the 2PN results lies
within the 1PN results, indicating that an inclusion of even higher PN
orders would refine our findings, but is unlikely to move the results in a different region.
This plot also demonstrates the importance of 2PN knowledge. The 2PN results show that dynamical tidal
effects are important at least up to a mass ratio of 3, in agreement with Fig.~\ref{isco}.
It is also intriguing that the two Taylor expanded alternative models TEOB-$\mu_{\rm DT}^{e}$ and TEOB-$A_{\rm DT}^{e}$ lead to very
different results at 1PN, but their factorized versions agree quite well. This demonstrates
that the factorization can remove some arbitrariness from the EOB resummation.

We further note from Fig.~\ref{dphi} that the effective--Love-number model TEOB-$k_{\rm eff}$ captures the effects well, 
in spite of the derivation of $k_{\rm eff}$ being based on Newtonian gravity and leading-order radiation reaction.
However, in hindsight this makes sense because (i) the model includes relativistic corrections to the tidal field with 2PN accuracy since the
effective--Love-number function enters through Eq.~(\ref{AAT2PN}), and (ii) the
relativistic redshift and frame-dragging effects tend to compensate each other (as explained in Sec.~\ref{framedrag}), thus leading to only a small shift of the resonance condition away from the Newtonian expectation.
Since the TEOB-$k_{\rm eff}$ model does not require an evolution of additional dynamical variables it is more convenient for generating a large bank of gravitational waveforms.
Furthermore, the Love number and f-mode frequency are linked by an approximately universal relation \cite{Chan:2014kua}, which can be used to reduce the parameter space for
the template bank. A universality of this kind
can also potentially be used as a test of general relativity, as discussed in Ref.~\cite{Yagi:2013bca}.

Previous studies have raised concerns regarding the measurement of tidal effects due to
the lack of knowledge of high-order terms in the PN approximation in the
point-mass sector \cite{Favata:2013rwa, Yagi:2013baa}. This is because, as discussed in Sec. III, Newtonian tidal effects enter formally at 5PN order, but the point-mass terms are only known to 4PN order, so that this lack of PN knowledge could lead to systematic errors in the measurements of tidal parameters. However, this issue arises only for PN-based analytical waveform models. It is mitigated in the EOB model since its point-mass version is resummed and calibrated to numerical relativity~\cite{Taracchini:2013rva} and thus effectively includes all the higher PN orders. For nonspinning binaries the systematic errors in the EOB model have been quantified and found to be small. Therefore, EOB-based measurements of tidal parameters for such systems are not expected to be contaminated by the large systematic errors found in \cite{Favata:2013rwa, Yagi:2013baa}. This issue also illustrates why synergetic approaches like EOB are important to obtain accurate waveforms over the full frequency range of current detectors.

\section{Conclusion}

In this paper we developed a general relativistic model
for dynamic tides based on a covariant effective action. While we focused our analysis on the quadrupolar f-mode oscillations of neutron stars, the results can readily be extended to more general cases and higher multipoles. We derived explicit results for all the contributions to this action both in the PN and the test-particle limit and discussed the physical effects encoded in these results. This served as the foundation for constructing, for the first time, an EOB Hamiltonian describing dynamical tidal effects both for generic orbits and specialized to circular orbits. In contrast to the line of work in
Refs.~\cite{Bini:2012gu, Bernuzzi:2014owa}, our TEOB model does not contain
poles at the light ring due to the choice of gauge we adopted,\footnote{A comparison of our TEOB model
and the models in Refs.~\cite{Bini:2012gu, Bernuzzi:2014owa} against numerical-relativity simulations can be found
in Refs.~\cite{Hinderer:2016eia,inprep2}.} but it still reproduces the
test-particle limit. Throughout these derivations we provided the relevant details of the calculations to make the paper self-contained and highlighted various subtleties. We then used the new tidal EOB Hamiltonian to show that dynamical tides are relevant both in the conservative dynamics and in the gravitational-wave phase and quantified the uncertainty in the model due to the lack of higher-order tidal PN information. Moreover, we devised a computationally more efficient yet approximate TEOB model where the dynamical tidal effects are encoded in an effective--Love-number function which we calculated. 

Our model is currently being implemented for gravitational-wave data
analysis and will aid in extracting the unique information on the
equation of state of neutron stars from upcoming observations with 
Advanced LIGO and Virgo. In a forthcoming paper
\cite{inprep2} we will study refinements of the EOB waveform model,
include dynamical higher multipoles, as well as the effects of
dynamical tides in the dissipative sector and hence in the waveform
amplitudes, and perform comparisons of the model against new highly 
accurate numerical-relativity simulations of neutron-star--black-hole binary systems.

\acknowledgments
We gratefully acknowledge useful discussions with Justin Vines and Francois Foucart. 
T. H. acknowledges support from NSF Grant No. PHY-1208881 and thanks the Max Planck 
Institute for Gravitational Physics in Potsdam for hospitality.

\appendix

\section{Point-mass parts of the effective-one-body potentials\label{PMEOB}}
Here we report the point-mass potentials from Ref.~\cite{Taracchini:2013rva} that enter our tidal EOB model. The potential $A$ is given by
\begin{widetext}

\be
A=\bar{\Delta}_u \left(\Delta _0 \nu +\nu 
   \log \left(\Delta_5 u^5+\Delta_4
   u^4+\Delta_3 u^3+\Delta_2
   u^2+\Delta_1 u+1\right)+1\right),
   \ee
with
\bes
\bea
\bar{\Delta}_u&=&a^2\left(u-\frac{1}{r_+}\right)\left(u-\frac{1}{r_-}\right),\\
r_\pm&=&\left(1\pm\sqrt{1-a^2}\right)\left(1-\nu K\right),\\
\Delta_5&=& \frac{(K \nu -1)^2}{\nu } \bigg[ \frac{64}{5} \nu  \log  (u)
  + \nu  \left(-\frac{1}{3} a^2
   \left(\Delta_1^3-3\Delta_1
  \Delta_2+3 \Delta _3\right)+\frac{\Delta_1^4-4 \Delta _1^2\Delta_2+4\Delta_1\Delta_3+2\Delta_2^2-4 \Delta_4}{2 K \nu  -2}\right. \nonumber \\
  &&\left.-\frac{\Delta_1^5-5\Delta_1^3
  \Delta_2+5\Delta_1^2 \Delta_3+5\Delta_1\Delta_2^2-5
  \Delta_2 \Delta_3-5\Delta_4\Delta_1}{5 (K \nu -1)^2}+\frac{2275 \pi^2}{512}+\frac{128 \gamma }{5}-\frac{4237}{60}+\frac{256
   \log (2)}{5}\right) \bigg] , \\
   \Delta_4&=&\frac{1}{96} \bigg[8 \left(6 a^2 \left(\Delta_1^2-2 \Delta_2\right) (K \nu -1)^2+3 \Delta_1^4+\Delta_1^3 (8-8 K \nu )-12 \Delta_1^2\Delta_2+12\Delta_1 (2
  \Delta_2 K \nu -2 \Delta_2+\Delta_3)\right)\nonumber\\
  &&+48\Delta_2^2-64 (K \nu -1) (3 \Delta_3-47 K \nu +47)-123 \pi ^2 (K\nu -1)^2\bigg],\\
   \Delta_3&=&-a^2\Delta_1 (K \nu -1)^2-\frac{\Delta_1^3}{3}+\Delta_1^2 (K \nu -1)+\Delta_1\Delta_2-2 (K \nu -1) (\Delta_2-K
   \nu +1),\\
   \Delta_2 &=& \frac{1}{2} \left(\Delta_1 (\Delta_1-4 K
   \nu +4)-2 a^2\Delta_0 (K \nu -1)^2\right),\\
   \Delta_1 &=&-2 (\Delta_0+K) (K \nu -1),\\
   \Delta_0&=&K (K \nu -2),
\eea
\ees
\end{widetext}
where $K$ is a calibration parameter tuned to numerical-relativity simulations whose value is given in Ref.~\cite{Taracchini:2013rva}. The potential $D_\text{\PM}$ is
\be
D_\text{\PM}=1+\log \bigg[1+\frac{6 \nu 
   G^2M^2}{r^2}+\frac{2 (26-3 \nu ) \nu  G^3M^3}{r^3}\bigg].
\ee
In all expressions above we use only the nonspinning limit where $a\to 0$. 
In our implementation, we evolve the
``tortoise'' radial momentum
\begin{equation}
p_{r*} = \frac{p_r}{\sqrt{D}} ,
\end{equation}
instead of $p_r$, and for the non-geodesic term we use
\begin{equation}
\frac{\mu^2_\text{\PM}}{\mu^2} = 2 \nu (4-3\nu) \frac{p_{r*}^4 G^2 M^2}{\mu^4 r^2} .
\end{equation}

\section{Equilibrium and adiabatic solutions\label{equilibrium}}

Equilibrium solutions are solutions for $Q^{ij}$ that are static in
the corotating frame and exist for circular orbits. These solutions
are obtained by solving for $Q^{ij}_{\rm equil}$ when setting to zero
the time derivatives of the equations of motion:
${\partial H_{\rm EOB}}/{\partial Q^{ij}}
\mid_{p_r=0}=0$,
${\partial H_{\rm EOB}}/{\partial P_{ij}}\mid_{p_r=0}=0$.
Here, we give the
specific solutions for the variables $(\alpha, \beta,\gamma)$ defined
in Eq.~(\ref{eq:corotatingQ}) for the case of our TEOB model, the
generalization to other tidal resummations can be derived from the above 
equilibrium equations.  When written out explicitly, the EOB tidal
potentials in the circular-orbit limit are 
\begin{widetext}
\bes
\bea
\frac{\mu^2_{\rm DT}}{\mu^2}&=&-\frac{3 G^2M X_2 (\alpha +\beta )}{\nu  r^4}
   \left(2-(1-c_1) \nu \right) \left(1+\frac{3
  G M}{r}\right)+\frac{6 \beta  G^2M
   X_2 \left(1+\frac{3 GM}{r}\right)}{\nu 
   r^4}\nonumber\\
   &&+\frac{2}{\mu}
   \left(1+\frac{3 GM
   X_1}{2 r}+\frac{27 G^2M^2 X_1}{8 r^2}\right) \bigg[\frac{3 \alpha
   ^2+\beta ^2+\gamma ^2}{2 \lambda }+\frac{1}{6}
   \lambda  \omega_f^2 \left(p_\alpha^2+3 p_\beta^2+3 p_\gamma^2\right)\bigg],\\
A_{\rm DT}&=&-\frac{3 GM X_2 (\alpha +\beta )}{\mu  r^3}
\bigg[1+\frac{5 G^2M^2 X_1 (33
   X_1-7)}{28 r^2}+\frac{GM ((1-c_1) \nu -2
   X_2)}{r}\bigg],\\
 f_{\rm DT}&=&-\frac{2 \sqrt{GM}   (\beta  p_\gamma -\gamma  p_\beta)}{\mu  r^{3/2}}\bigg[1-\frac{GM (\nu +3 X_2)}{2 r}-\frac{G^2M^2 \left(\nu ^2+27
   \nu -6 \nu  X_2+9 X_2\right)}{8
   r^2}\bigg].
\eea     
\ees
\end{widetext}
From $\dot \alpha=0$ we obtain $p_\alpha^{\rm equil}=0$, and from both
$\dot \beta=0=\dot p_\gamma$ we find $\gamma^{\rm equil}=0=p_\beta^{\rm equil}$.
To proceed further requires either
numerically solving the equations $0=\partial H_{\rm eff}/\partial
\alpha=\partial H_{\rm eff}/\partial \beta=\partial H_{\rm
  eff}/\partial p_\gamma$ for $\alpha, \beta, p_\gamma$ or making a
perturbative expansion by linearizing in the tidal terms. The results
of this can be obtained explicitly with Mathematica, but are
not particularly illuminating. Note that when doing a PN expansion one
cannot brute-force expand the full EOB solutions for $r\to \infty$
since this would also PN-expand the ``Newtonian'' dependence
$1/[1-\omega_f^2/(4\Omega^2)]$ (or with EOB involving $p_\phi$ rather
than $\Omega$). When solving the equations of motion iteratively for
$\beta=\beta_{\rm Newt}+\beta_{\rm PN}$ etc. we obtain
\begin{widetext}
\bes
\label{eq:equil}
\bea
\beta^{\rm equil}&=&\frac{3 \lambda  G M  X_2}{2 r^3
   (1-W)}-\frac{3 \lambda  G^2M^2 X_2
   (2 \nu W +(1-W) (X_2-3))}{4 r^4
   (1-W)^2},\\
    p_\gamma^{\rm equil}&=&\frac{3 \sqrt{G M} W X_2}{4 r^{3/2}
   (1-W)}-\frac{3 (GM)^{3/2} W X_2 \left[\nu(1 +W)
 +(X_2-3)(1-W)\right]}{8 r^{5/2}
   (1-W)^2},\\
    \alpha^{\rm equil}&=&\frac{\lambda  G M X_2}{2 r^3}-\frac{\lambda (G M)^2
   (X_2-7) X_2}{4 r^4},
   \eea
   \ees
where $W={4GM}/(r^3 \omega_f^2)$. The adiabatic limit is obtained for  $W\to 0$ or $\omega_f^2\gg \Omega^2\sim GM/r^3$ in Eqs.~(\ref{eq:equil}) and leads to
\be
\beta^{\rm AT}=\frac{3 \lambda G M X_2}{2 r^3}-\frac{3 \lambda  G^2M^2
   (X_2-3) X_2}{4 r^4}, \ \ \ \ \ p_\gamma^{\rm AT}=0, 
   \ee
   and $ \alpha^{\rm AT}=\alpha^{\rm equil}$. For the initial conditions we use the circular-orbit solution for $P_\phi$ (valid again for our TEOB model, but the generalization to other models simply requires setting to zero $f_{\rm DT}$ and $\mu^2_{\rm DT}$ or $A_{\rm DT}$),
\be
\frac{p_\phi^2 \!\mid_{\rm circ} }{\mu^2} =-\frac{2 r^3 A f_{\rm DT}^\prime
   \sqrt{\left(2 A-r A^\prime\right) \left(r
   ({\tilde\mu}^2_{\rm DT})^\prime+2 {\tilde\mu}^2_{\rm DT}+2\right)+r^2
  \left( f_{\rm DT}^\prime\right)^2}}{\left(r A^\prime-2
   A\right)^2}-\frac{ r^3
   \left(({\tilde\mu}^2_{\rm DT}+1) A^\prime+A
   (\tilde\mu^2_{\rm DT})^\prime\right)}{r A^\prime-2 A}+\frac{2
   r^4 A \left(f_{\rm DT}^\prime\right)^2}{\left(r
   A^\prime-2 A\right)^2} .
\ee
\end{widetext}
Here $\tilde{\mu}_\text{DT} = \mu_\text{DT} / \mu$, primes denote derivatives with respect to $r$, and all tidal potentials are evaluated for the equilibrium solutions computed numerically as described above. We augment the nontrivial solutions for $p_\phi$, $\alpha$, $\beta$, $p_\gamma$ by the initial value for $p_r$. This is computed from numerically solving for $p_r$ from 
\be
\left. \frac{\dot{E}\, \left(\partial^2 H_{\rm EOB}/\partial r\partial p_\phi \right)}{\left(\partial H_{\rm EOB}/\partial p_\phi\right)\left(\partial^2 H_{\rm EOB}/\partial r^2 \right)}\right|_{\rm circ}=-\frac{\partial H_{\rm EOB}}{\partial p_r}.
\ee

\section{The oscillator Hamiltonian and the mapping from post-Newtonian to effective-one-body Hamiltonians\label{massshift}}

In this appendix we discuss some subtleties in the identifications of tidal terms in the EOB model that arise when starting from the structures in the PN Hamiltonian instead of basing the construction on the test-particle limit. Whereas in the test-particle case we can obtain additional information from the mass-shell constraint (see Sec.~\ref{TPEOB}), this information is not readily available in the PN limit where our explicit results are limited to the Hamiltonian. Below we discuss the consequences of this imbalance in the source of information in the two limits. We start by outlining several arguments for adding tidal terms into the various EOB functions similar to those for the test-particle limit. While for the interaction terms both the Newtonian limit and test-particle expectations lead to consistent identifications, the oscillator terms give rise to a discrepancy that we discuss and resolve.  

The structure of the leading-order PN-tidal corrections can be
identified in a similar manner as discussed in the context of the
test-particle, namely by counting the power of momenta in each
term. First, we note that based on our assumptions, the effective metric is
independent of the canonical momentum. As a consequence, 
the structure of Eq.~(\ref{Heff}) dictates that (i) interactions that are
linear in $p_i$ must be incorporated in the potential $\beta^i$, (ii) terms
quadratic in $p_i$ should appear in $\gamma^{ij}_\text{eff}$, and (iii) 
terms independent of the momentum must be in $A$.  Remaining terms of cubic and higher order in
$p_i$ are then collected into $\mu_\text{NG}$. Following this
reasoning we deduce that the Newtonian interaction term in
Eq.~(\ref{H1PN}), which is independent of the momenta, belongs
to $A$. This agrees with the result of applying similar arguments in the
test-particle case to the second term on the right-hand side of Eq.~(\ref{HEQTP}). However,
this consistency between PN and test-particle--limit identifications fails for the oscillator piece.

We have deduced in Sec. IV B that in the test-particle limit the pure
oscillator Hamiltonian enters the EOB functions through the
nongeodesic term $\mu^2_\text{NG}$. On the other hand, following the
reasoning for the EOB identification of PN corrections we note that in
the Newtonian limit the oscillator Hamiltonian (\ref{Ho}) with $z_1 =
1$ does not depend on the canonical linear momentum.  Following the
classification of terms by powers of momenta, it should therefore be
included in $A$ instead of $\mu_\text{NG}$. This discrepancy is due to
the additional information from the $p_0$-dependence in the mass-shell
constraint (\ref{mshell}), which is available in the
test-particle limit, but not in the PN Hamiltonian. This means that
the test-particle limit gives a more refined picture in this case, so
we include the oscillator part in $\mu_\text{NG}^2$ here.

The freedom in making the identifications between PN tidal terms and
the EOB Hamiltonian can also be exploited to devise different
mappings. For instance, adopting the convention that
momentum-independent terms should be included in $\mu^2_\text{DT}$ in
the PN case would shift the disagreement with the test-particle
mass-shell constraint to the $H_\text{QE}$ contributions. However, it
is important to stress that these ambiguities have no physical
consequences and are merely a result of incomplete information within
the different approximation schemes. In particular, note that PN
information enters in the oscillator Hamiltonian~(\ref{Ho}) only
through the redshift $z$. An accurate prediction for the value of the
redshift beyond the PN expressions is provided by the EOB point-mass
Hamiltonian through
\begin{equation}\label{redshiftHEOB}
z_A = \frac{\partial H_\text{EOB}^\text{\PM}}{\partial m_A} .
\end{equation}
Since $H_\text{EOB}^\text{\PM}$ has been calibrated to numerical-relativity 
simulations for circular orbits, this formula gives the redshift $z_A$ to high accuracy and could be used to improve the
resummation of the pure oscillator terms in any of the EOB potentials.

Finally, we point out another interesting possibility for a resummation. The Hamiltonian (\ref{Ho}) together
with Eq.~(\ref{redshiftHEOB}) is the first term in a Taylor expansion in the
mass $m_1$. The most elegant way to include the oscillator terms is therefore
a shift of the mass $m_1$ given by
\begin{equation}
m_1 \rightarrow m_1 + \lambda \omega_f^2 P_{ij} P_{ij}
 + \frac{1}{4 \lambda} Q^{ij} Q^{ij} ,
\end{equation}
in $H_\text{EOB}^\text{\PM}$. This automatically makes the oscillatory dynamics
as accurate as $H_\text{EOB}^\text{\PM}$. However, it implies that dynamical
terms are introduced in the energy map (\ref{HEOB}) as well. Since the tidal effects
are small, we do not further explore this proposal
here, but it is worth to point out that such a modification of the energy map would lead to a noticeable structural simplification.

\section{Canonical transformations and the pole at the light ring\label{POLE}} 
In this appendix we consider the effect of using a canonical
  transformation to specialize the test-particle--limit tidal
  Hamiltonian~(\ref{HTPLexpand}) to circular orbits. The general
  method was explained in Sec.~\ref{qcirc} and here, we only provide
  an illustrative example for one of the terms in the Hamiltonian.
  This serves to clarify the statements made in Ref.~\cite{Akcay:2012ea}
  that the pole at the light ring comes from a particular gauge choice 
and it can be eliminated through a
  canonical transformation. In other words, the light-ring pole
    should be interpreted as a coordinate singularity in the phase
    space.

In the test-particle limit, the generator $g_f$ from Eq.~(\ref{gf}) leads to the transformation
\begin{align}\label{gTPL}
&\{ H_\text{\PM}^\text{TPL}, g_f \}
= - f(\vct{r},\vct{p},Q^{ij},P_{ij}) \frac{r \dot{p}_r}{\mu} + \Order(p_r) \\
&= \frac{f(\vct{r},\vct{p},Q^{ij},P_{ij}) \mu}{H_\text{\PM}^\text{TPL}}
  \left[- \frac{\vct{p}^2}{\mu^2} (1-3u) + u \right] + \Order(p_r).
\end{align}
We next use this relation to eliminate $\vct{p}^2$ from the tidal part of the test-particle--limit Hamiltonian~(\ref{HTPLexpand}) in favor of its circular-orbit value as a function of $u$ given by Eq.~(\ref{circularTPL}),
\begin{equation}\label{circularTPLappend}
\vct{p}^2 = \frac{\mu^2 u}{1 - 3 u} + \Order(p_r) ,
\end{equation} 
which exhibits the pole at the light ring. Note that the occurrences of $\vct{p}^2$ in the tidal part of Eq.~(\ref{HTPLexpand}) enter both through 
 the overall prefactor $z_\text{TPL}$, determined from Eqs.~(\ref{zTPL}) and (\ref{HTPL1}), and through the interaction term in Eq.~(\ref{Htest}). This can be analyzed by working with the binomial expansion
 \be
 \label{zTPLbinom}
z_\text{TPL}=\sqrt{A_\text{TPL}}\sum_{n=0}^\infty \frac{(-1)^n(2n-1)!!}{2^n n!}\left(\frac{\vct{p}^2}{\mu^2}\right)^{n} ,
\ee
where we used $\vct{p}_e^2 = \vct{p}^2$ following from $p_r = 0$.
For example, consider the term involving the second combination in (\ref{HEQTP}), which enters into the Hamiltonian~(\ref{HTPLexpand}) in the form
\begin{align}
H_\text{EQ,$p^2$}^\text{TPL} &= - \frac{3 G M}{2 \mu^2 r^3} z_\text{TPL} Q^{ij} n^i n^j \vct{p}^2,\label{HTPLfullp2}\\
&= \frac{3 G M}{2 r^3}Q^{ij} n^i n^j \sqrt{A_\text{TPL}} \left[-\frac{\vct{p}^2}{\mu^2} + \frac{\vct{p}^4}{2 \mu^4} - \Order(\vct{p}^6)\right] , \label{HTPLp2}
\end{align}
where in Eq.~(\ref{HTPLp2}) we explicitly consider only the first two terms in the expansion of $z_{\rm TPL}$ from (\ref{zTPLbinom}). 
If we use
\begin{equation}
f_0 = \frac{ H_\text{\PM}^\text{TPL}}{1 - 3 u} \frac{3 G M}{2 \mu r^3} Q^{ij} n^i n^j
  \sqrt{A_\text{TPL}} \left[- 1 + \frac{\vct{p}^2}{2\mu^2} - \Order(\vct{p}^4)\right] ,
\end{equation}
we can eliminate the first occurrence of $\vct{p}^2$ from the transformed Hamiltonian
\begin{multline}\label{circularExampleTPL}
H_\text{EQ,$p^2$}^\text{TPL} + \{ H_\text{\PM}^\text{TPL}, g_{f_0} \} =
   \frac{3 G M}{2  r^3}Q^{ij} n^i n^j \\
     \times \sqrt{A_\text{TPL}} \left[-\frac{ u}{1 - 3 u} + \frac{u}{2(1 - 3 u)} \frac{\vct{p}^2}{\mu^2} - \Order(\vct{p}^4) \right] .
\end{multline}
To remove the remaining dependence on $\vct{p}^2$ we apply a second transformation with
\begin{equation}
f_1 = \frac{H_\text{\PM}^\text{TPL}}{1 - 3 u} \frac{3 G M}{2 \mu r^3} Q^{ij} n^i n^j \sqrt{A_\text{TPL}}
    \left[ \frac{u}{2(1 - 3 u)} - \Order(\vct{p}^2) \right] ,
\end{equation}
and obtain
\begin{multline}\label{circularExampleTPL1}
H_\text{EQ,$p^2$}^\text{TPL} + \{ H_\text{\PM}^\text{TPL}, g_{f_0} \} + \{ H_\text{\PM}^\text{TPL}, g_{f_1} \} =
  \frac{3 G M}{2  r^3}Q^{ij} n^i n^j \\
    \times \sqrt{A_\text{TPL}} \left[-\frac{ u}{1 - 3 u} + \frac{u^2}{2(1-3u)^2} - \Order(\vct{p}^2) \right].
\end{multline}
Repeating this procedure and summing the series for which we only exhibited the first two terms leads to
\bea
H_\text{EQ,$p^2$}^\text{TPL~circ}&=&H_\text{EQ,$p^2$}^\text{TPL} + \left\{ H_\text{\PM}^\text{TPL}, g_f \right\}\\
&=& - \frac{3 G M}{2 r^3}Q^{ij} n^i n^j \frac{ u}{\sqrt{1 - 3 u}} ,
\eea
where $g_f = \sum_n g_{f_n}$ or $f = \sum_n f_n$.
This rigorously demonstrates that simply substituting Eq.~(\ref{circularTPLappend}) into Eqs.~(\ref{zTPL}) and (\ref{HTPLfullp2}) is a valid procedure to specialize to circular orbits and introduces an explicit pole at the light ring $u=1/3$.

Furthermore, as first noticed in Ref.~\cite{Akcay:2012ea}, the transformation 
outlined above introduces a coordinate singularity in the phase space at the light 
ring. Here, we made it explicit that the singularity is produced by the poles in the 
generator of the canonical transformation $g_f$.
Nevertheless, the presence of 
poles is not problematic as long as the 
light ring is not reached. An important observation is that the method of 
the canonical transformation works in both ways, i.e., one can also remove an
explicit pole at the light ring
by replacing it with a function of $\vct{p}^2$ using Eq.~(\ref{circularTPLappend}).
In the explicit example given above, this corresponds to performing the inverse
canonical transformation generated by minus $g_f$.



%

\end{document}